\PassOptionsToPackage{table}{xcolor}
\documentclass[conference]{IEEEtran}
\IEEEoverridecommandlockouts
\usepackage{cite}
\usepackage{amsmath,amssymb,amsfonts}
\usepackage{algorithmic}
\usepackage{graphicx}
\usepackage{textcomp}
\usepackage{times}
\def\BibTeX{{\rm B\kern-.05em{\sc i\kern-.025em b}\kern-.08em
    T\kern-.1667em\lower.7ex\hbox{E}\kern-.125emX}}

\usepackage{orcidlink}
\usepackage{multirow}
\usepackage{graphicx}
\usepackage{hyperref}
\usepackage{caption}
\usepackage{rotating}
\usepackage{adjustbox}
\usepackage[inline]{enumitem}

\makeatletter
\renewcommand\IEEEkeywordsname{Keywords}
\makeatother
\usepackage{xpatch}
\xpatchcmd\IEEEkeywords{---}{-}{}{}

\makeatletter
\renewcommand{\fnum@figure}{Figure~\thefigure}
\makeatother

\begin{document}

\title{\bfseries\Large Designing at 1:1 Scale on Wall-Sized Displays Using Existing UI Design Tools
\\
}

\author{\IEEEauthorblockN{1\textsuperscript{st} Lou Schwartz}
\IEEEauthorblockA{%\textit{dept. name of organization (of Aff.)} \\
\textit{Luxembourg Institute of Science and Technology (LIST)}\\
Esch-sur-Alzette, Luxembourg \\
lou.schwartz@list.lu~\orcidlink{0000-0002-8645-5326}}
\and
\IEEEauthorblockN{2\textsuperscript{nd} Mohammad Ghoniem}
\IEEEauthorblockA{%\textit{dept. name of organization (of Aff.)} \\
\textit{Luxembourg Institute of Science and Technology (LIST)}\\
Esch-sur-Alzette, Luxembourg \\
mohammad.ghoniem@list.lu~\orcidlink{0000-0001-6745-3651}}
\and
\IEEEauthorblockN{3\textsuperscript{rd} Valérie Maquil}
\IEEEauthorblockA{%\textit{dept. name of organization (of Aff.)} \\
\textit{Luxembourg Institute of Science and Technology (LIST)}\\
Esch-sur-Alzette, Luxembourg \\
valerie.maquil@list.lu~\orcidlink{0000-0002-0198-3729}}
\and
\IEEEauthorblockN{4\textsuperscript{th} Adrien Coppens}
\IEEEauthorblockA{%\textit{dept. name of organization (of Aff.)} \\
\textit{Luxembourg Institute of Science and Technology (LIST)}\\
Esch-sur-Alzette, Luxembourg \\
adrien.coppens@list.lu~\orcidlink{0000-0002-2841-6708}}
\and
\IEEEauthorblockN{5\textsuperscript{th} Johannes Hermen}
\IEEEauthorblockA{%\textit{dept. name of organization (of Aff.)} \\
\textit{Luxembourg Institute of Science and Technology (LIST)}\\
Esch-sur-Alzette, Luxembourg \\
johannes.hermen@list.lu~\orcidlink{0000-0002-1198-5701}}
}

% \author{
%     \IEEEauthorblockN{Lou\ Schwartz\,\orcidlink{0000-0002-8645-5326}}
%     \IEEEauthorblockN{Mohammad\ Ghoniem\,\orcidlink{0000-0001-6745-3651}}
%     \IEEEauthorblockN{Valérie\ Maquil\,\orcidlink{0000-0002-0198-3729}}
%     \IEEEauthorblockN{Adrien\ Coppens\,\orcidlink{0000-0002-2841-6708}}
%     \IEEEauthorblockN{Johannes\ Hermen\,\orcidlink{0000-0002-1198-5701}}
%     \IEEEauthorblockA{%
%         Luxembourg Institute of Science and Technology (LIST)\\
%         5, avenue des Hauts-Fourneaux, L-4362 Esch-sur-Alzette, Luxembourg\\
%         % Diferring from IEEE ("Email"), IARIA requires "e-mail":
%         %e-mail: \texttt{c.neumann@oth-aw.de}
%         % Multiple authors and their e-mail addresses:
%         e-mail: \small{\tt$\lbrace$lou.schwartz\,|\,mohammad.ghoniem\,|\,valerie.maquil\,|\,adrien.coppens\,|\,johannes.hermen|\,$\rbrace$@list.lu}
%     }
% }

\maketitle

\begin{abstract}
Wall-Sized Displays have spatial characteristics that are difficult to address during user interface design. The design at scale 1:1 could be part of the solution. In this paper, we present the results of two user studies and one technology review, exploring the usability of popular, desktop-optimized prototyping tools, for designing at scale on Wall-Sized Displays. We considered two wall-sized display setups, and three different interaction methods: touch, a keyboard equipped with a touchpad, and a tablet. We observed that designing at scale 1:1 was appreciated. Tablet-based interaction proved to be the most comfortable interaction method, and a mix of interaction modalities is promising. In addition, care must be given to the surrounding environment, such as furniture. We propose twelve design guidelines for a design tool dedicated to this specific context. 
Overall, existing user interface design tools do not yet fully support design on and for wall-sized displays and require further considerations in terms of placement of user interface elements and the provision of additional features.
\end{abstract}


\section{Introduction}

%\color{red}
This paper extends a previous study on the usability of Figma, a popular user interface (UI) design tool, for Wall-Sized Displays (WSDs) at the CENTRIC 2024 conference~\cite{centricSchwartz}. 
WSDs are increasingly used in public spaces to provide general or contextual information, provide entertainment, or for artistic purposes~\cite{parker2020foundations}\cite{finke2008lessons}.
WSDs are also applied in many research areas. Traffic management~\cite{prouzeau2016towards} and automotive design~\cite{buxton2000large} benefit from their large display area. They also support data browsing and manipulation ~\cite{liu2016shared}\cite{ wigdor2009wespace}, and are crucial for visualizing and interacting with vast amounts of data in natural sciences like physics, astronomy, chemistry, and biology~\cite{rajabiyazdi2015understanding}\cite{pietriga2016exploratory}. In the medical field, WSDs aid interdepartmental communication~\cite{simonsen2020infrastructuring}, optimize surgery room scheduling, and improve team transitions~\cite{rambourg2018welcome}\cite{thomas2017echo}. They are also used for scheduling activities, such as conference organization~\cite{doshi2017stickyschedule}, and support collaborative design tasks~\cite{wall2020decision}, including architectural design reviews~\cite{kubicki2019assessment}.
%\color{black}

WSDs are also referred to as vertical Large Interactive Displays (LIDs) or Large High-Resolution Displays (LHRDs). However, the notion of `large' is typically not precisely defined and can be subjective~\cite{lischke2020challenges}. Belkacem et al. defined an LHRD as a display that ``\textit{creates a coherent physical view space that is at least of the size of the human body and exhibits a significantly higher resolution than a conventional display}''~\cite{belkacem}.
According to Chen et al., WSDs improve user performance and satisfaction for tasks, such as model design, analysis, and visual data mining~\cite{chen2021use}. 
But, these new ways of viewing, collaborating and interacting differ from desktop and smartphone applications~\cite{lischke2020challenges}, because of their size, their resolution, the collaboration they foster, and the so-called \textit{natural} interactions they enable, mainly through touch and gestures~\cite{ardito2015interaction}. 
%\color{red}
WSDs vary in terms of visualization technology, display setup (size, orientation), interaction modality, application objectives (productivity, entertainment, social interaction, games, and advertising) and location (city, office, education, conference, third place and cultural site)~\cite{ardito2015interaction}. %\color{black}

%\color{red}
Based on previous work, 
we enumerate nine challenges raised by WSDs, each needing further research~\cite{lischke2020challenges}\cite{belkacem}\cite{schwartz2023defis}: 
\begin{enumerate*}[label=\arabic*)]
    \item \emph{More interactions}: ``natural'' interactions with a mix of interaction methods~\cite{lischke2020challenges}\cite{schwartz2023defis}. 
    \item \emph{More users}: and improvements regarding how they collaborate~\cite{lischke2020challenges}\cite{belkacem}\cite{schwartz2023defis}.
    \item \emph{More space (around)}: users' movements and interactions in the space and the management of the inherent fatigue~\cite{belkacem}\cite{schwartz2023defis}. 
    \item \emph{Different usage durations}: for example, collaborative decision-making tasks require long sessions spread over time~\cite{schwartz2023defis}, but for public displays, the duration of use is often very short~\cite{lischke2020challenges}. 
    \item \emph{More complex content}: complex data representations, large quantities of complex data needed by experts~\cite{lischke2020challenges}\cite{belkacem}\cite{schwartz2023defis}.%%, with many questions about what information to display, how and where), 
    \item \emph{More devices}: WSDs are often composed of several displays, but are also often supplemented by other devices (e.g., supplementary displays or sensors)~\cite{lischke2020challenges}\cite{belkacem}\cite{schwartz2023defis}.
    \item \emph{Variable screen space and pixel count}: displaying more data simultaneously requires content to be designed differently to distinguish between and manage the visible, accessible and useful parts of the display~\cite{lischke2020challenges}\cite{belkacem}\cite{schwartz2023defis}.
    \item \emph{Support for designers}: tools or methods for designing and testing are needed, as well as guidelines and best practices~\cite{lischke2020challenges}\cite{schwartz2023defis}, and
    \item \emph{Other concerns} arise like accessibility, compatibility, and portability from one WSD to another, and workflow management~\cite{schwartz2023defis}.
\end{enumerate*}
 \color{black}

Supporting designers with the right tools and methods to design applications for WSDs is hence an open challenge.
%\color{black}\color{red} 
Similar questions have also been raised regarding the design of data visualization interfaces for WSDs, specifically~\cite{belkacem}\cite{sinaei2024elephant}. Overall, the challenge is three-fold: the difficulty of scaling visual elements, e.g., text, the limited availability of design software, and the lack of widely adopted design guidelines. %\color{black}

In this paper, we address the \textit{designer support} challenge, i.e., the need for design and testing tools and methods~\cite{lischke2020challenges}\cite{schwartz2023defis}. 
We look into the design of a UI prototype in WSD environments at 1:1 scale and seek to assess the suitability of existing design tools for this purpose. 
By `UI prototyping', we mean the prototyping of the interface, functionalities, screen layouts and behaviors at the mid-fidelity level.
We focus on using popular UI design tools, such as Figma~\cite{figma} and Miro~\cite{Miro}, to prototype UIs in WSD environments at 1:1 scale. We have no conflict of interest with any of them.

% distance de l'écran + résolution --> lisibilité du texte
% déplacements ?

In the rest of this paper, Section~\ref{sec:sota} presents related work on methods and tools for prototyping for WSDs. Our research approach is detailed in Section~\ref{sec:RQ}. Section~\ref{sec:XP1} describes a first user study using Figma to design for WSD at 1:1 scale. This allowed us to identify the required features to support design at 1:1 scale for WSDs. Section~\ref{sec:benchmark} compares existing design tools based on these features. Section~\ref{sec:XP2} presents a second user study, diving deeper into the use of another design tool (Miro) with different WSD setups and interaction means. Our observations are discussed in Section~\ref{sec:discussion} and used to draw implications for design, and %guidelines, requirements and suggestions 
for tooling improvements. The section also discusses the limitations of both user studies. Finally, Section~\ref{sec:conclusion} holds a general conclusion.

%%In this paper, we make the focus on the UI design for WSDs, that means the design of the UI through mockups or prototyping.
%In this paper, we focus on the design of a user interface (UI) for a mid-fidelity prototype and seek to understand whether an existing online design tool can be used to prototype in the WSD environment at 1:1 scale. We focus on interface functionality and screen layouts and behaviour, summed-up in the term 'UI prototyping', at the mid-fidelity level.

%\section{Related work \textbar{} Methods}
%\lipsum[23]
\section{Related work}
\label{sec:sota}

Many tools and methods have been proposed to support design for WSDs. %\color{red} 
Below we discuss paper prototyping, prototype development and mixed mockup techniques. We also cover briefly interaction techniques used in WSD environments.

\subsection{Paper prototyping}
Paper prototyping is a popular, validated and simple method of mocking up systems before programming~\cite{snyder2003paper}. %\color{black}
% The value of this method %\textbf{Paper prototyping}\cite{snyder2003paper} 
It allows designers to explore, communicate and evaluate early interface designs with end-users or within the design team. A designer typically plays the role of the computer to simulate the behavior of the system by changing the pieces of paper shown to the participants. Numerous studies have used paper prototyping to design applications on a WSD (e.g.,%\color{red}
~\cite{doshi2017stickyschedule}%\color{black}
\cite{bailey2008adapting}\cite{avellino2017camray}%\color{red}
\cite{fallman2005designing}).

Bailey et al. used this method to prototype a multi-device environments (MDE) involving personal devices (e.g., computers or tablet), and a wide screen to share data and views~\cite{bailey2008adapting}. They observed that A4 paper lacks accuracy for large displays due to scaling issues. Indeed, the size hampers reading text at a distance. 
WSDs require sheets of paper larger than A4, and text size should be adapted to the screen size.
However, paper prototyping does not allow checking whether the user can quickly and easily detect where information is displayed and whether changes in the displayed content would be noticed, as the user can follow the facilitator's gaze direction and movements and see where she places the pieces of paper. 

Paper prototyping can also be used to define the screen arrangements to compose the WSD~\cite{lischke2016screen}. 

However, paper prototyping is mainly used for UI prototyping on WSDs.%\color{black}

%\color{red} 
\subsection{Functional prototyping}%\color{black}
Mid-fidelity prototype development is also a common practice~\cite{chegini2017interaction}.
%\color{red}
Indeed, in the following papers, for example, there are no indication on how the applications were designed, but a developed prototype is used for the studies within them~\cite{prouzeau2016towards}\cite{liu2016shared}\cite{wigdor2009wespace}\cite{wall2020decision}\cite{langner2018multiple}\cite{bezerianos2006mnemonic}. 

Some systems have been developed to support prototyping, among others, for WSDs. For instance, \textit{DEBORAh}, is a front-end web-based software layer that supports the orchestration of interactive spaces combining multiple devices and screens, including WSDs~\cite{vandenabeele2022deborah}. Additionally, \textit{jBricks}, a Java-based toolkit, enables the exploratory prototyping of interaction techniques and rapid development of post-WIMP applications for WSDs, particularly for tiled-displays~\cite{pietriga2011rapid}.%\color{black}

%\color{red} 
\subsection{Mixed mockup techniques}%\color{black}
Finally, mixed mockup techniques, e.g., \textit{Mini-studio}~\cite{kim2016ministudio}, consist in using a physical paper model of the system augmented with projected content. They are mainly used to prototype ubiquitous computing systems, but can also be used to mock up WSDs. \textit{SketchStudio} is another example, which combines 2D devices with 3D characters, resulting in a 2.5D animated scenario design tool for rapid prototyping of systems involving multiple users and multiple components or devices~\cite{kim2018sketchstudio}.
Such methods and tools offer the advantage of allowing the interactions around the WSDs to be tested. 
However, they are not accurate enough for a prototype of the screen layout and content, especially in contexts where large amounts of data and high resolution are required~\cite{belkacem}. 

Overall, the prototyping method is frequently used for designing WSD applications, but the way the design was conceived is usually not described. Among the few occurrences where the design process is documented, paper prototyping is the most commonly used method. We did not find any studies covering the design of a UI prototype on a WSD at 1:1 scale. 

%\color{red}
\subsection{Interaction techniques for WSD environments}
Various interaction modalities have been used with WSDs, ranging from classic mouse and keyboard to more advanced types of interaction leveraging touch, gaze, mid-air gestures, proxemics, handheld and wearable devices, and tangibles~\cite{belkacem}. 

Mouse and keyboard interaction is readily available in virtually all interface design tools used in common desktop environments. Users may work from a distance while being seated. The physical keyboard is very convenient for text entry, but text may lack legibility at a distance. Virtual keyboards have also been studied in the context of mobile devices~\cite{textMobile}, horizontal tabletops~\cite{textTabletops} and virtual environments~\cite{textVR}. They can be spawned on vertical display surfaces too, but hardly any work investigates their usability in this context. While mouse interaction is faster and more accurate than touch~\cite{jakobsen2016negotiating}, the mouse cursor is easily lost and clutching may be tedious when navigating across very large scenes.

The cursor can also be attached to other pointing devices, be it a touchpad or a handheld device such as remote controllers, or flysticks, or head-mounted displays through eye tracking. 
Handheld display devices, such as smartphones or tablets, can also be turned into pointing devices by attaching sensors to them, or by using them as touchpads to interact with the WSD. 

Direct touch can also be used on touch-enabled WSDs. On the one hand, touch is an intuitive interaction modality that exhibits high user performance and acceptance scores~\cite{touchWSD}. On the other hand, users often have to move across the WSD, or step up close or away from the WSD, to look at details, or get the big picture, respectively~\cite{fatigueWSD}. They may also have a hard time reaching the upper or the lower part of the WSD~\cite{prouzeau2017evaluating}.

Overall, many advanced interaction techniques may generate fatigue or muscle strain, e.g., gorilla arm~\cite{fatigueWSD}, when used for a long time. They also differ significantly from widespread mouse and keyboard interaction, in terms of accuracy, speed, and their hedonic value~\cite{touchWSD}, and may be less suited for certain tasks such as text entry. A more detailed discussion of these interaction techniques for WSDs can be found in the literature~\cite{belkacem}.

In this work, we evaluate the suitability of popular interface design tools, originally meant for the desktop, to design UIs for WSD environments at the 1:1 scale. Most design tools were not developed for WSD, nor optimized for advanced interaction modalities available in WSD environments.

%\color{red}

\section{Research approach}\label{sec:RQ}
As noted by Lischke et al., when it comes to WSDs, it ``\textit{is often not possible to prototype in the original size}''~\cite{lischke2020challenges}. Unlike using a desktop computer to design UIs for WSDs, prototyping in real size directly onto the targeted display could reduce complexity, give a sense of scale, and ensure that the target resolution is correctly achieved and exploited. It could also allow designers to check that the UI is visible at all distances and from all angles~\cite{belkacem}\cite{schwartz2023defis}. Hence, our interest in prototyping on a WSD is based on supporting the design in real size.

\subsection{Research questions}
As part of understanding how to support the design in 1:1 scale for WSDs, we focus on three main research questions:

\begin{enumerate}[itemindent=1em,label=\textbf{RQ\arabic*:}]
    \item \textbf{Can a desktop optimized tool be used in a WSD environment to design at 1:1 scale?}
    \item \textbf{What would be the best interaction modality during the design at 1:1 scale on a WSD ?} And what are the main issues raised by each interaction modality?
    \item \textbf{What are the main features needed to support the design at 1:1 scale on a WSD ?}
\end{enumerate}

This research aims to extract initial guidelines for designing at 1:1 scale \emph{on} and \emph{for} WSDs. To answer these questions, we have set up two exploratory user studies and performed a comparative analysis of existing design tools. %feature matrix analysis.
The first user study involved one participant who consecutively interacted through three distinct interaction methods (a keyboard and its embedded touchpad, direct touch on the WSD, or a connected tablet) and with two WSDs to understand whether and how a desktop optimized tool, Figma, could be used in a WSD environment to design at 1:1 scale. 
The outcomes of the first user study pointed us to important features, as a basis for doing a technology review and comparative analysis, and selecting a better tool, i.e., Miro, for the next user study.
In the second user study, we observed the use of Miro by two participants designing at 1:1 scale on two WSDs with the same three interaction modalities. % (a keyboard and its embedded touchpad, direct touch on the WSD, or a connected tablet).

\subsection{Task}
\begin{figure*}[htbp]
   \centering
  \includegraphics[width=0.99\textwidth]{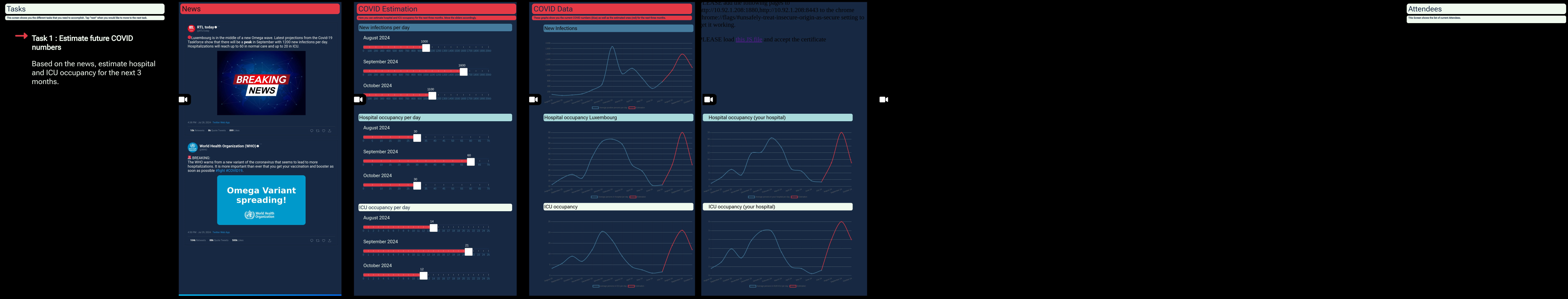}
  \caption{Screenshot of the prototype to replicate, for more information see~\cite{maquil2023establishing} and~\cite{annex1}\cite{annex2}.
  } 
  \label{fig:task}
\end{figure*}

In both user studies, the task consisted in using a desktop optimized software (Figma or Miro) %\color{black} 
to reproduce a previously developed UI~\cite{maquil2023establishing} %\color{red}
and adapted %\color{black} 
for both WSDs as shown in Figure~\ref{fig:task}, more details can be found online~\cite{annex1}\cite{annex2}. 
Before the test session, the participants discovered the design tool on a desktop computer for two hours, and the mock-up they had to reproduce.

This UI was chosen because it comprises a variety of UI elements (text, sliders, graph, a social media feed). Reproducing an existing UI ensures that it is feasible, well adapted to the WSD environment, and allows the observation to be focused on the design software manipulation rather than the process of creating a new design.

%\color{red}
\subsection{System}%\color{black} 
The system consisted of a touch-enabled WSD displaying the design tool (either Figma or Miro) in a web browser window (Google Chrome~\cite{chrome}) in full-screen mode. %\color{black}

\begin{figure}[ht]
  \centering
  \includegraphics[width=0.48\textwidth]{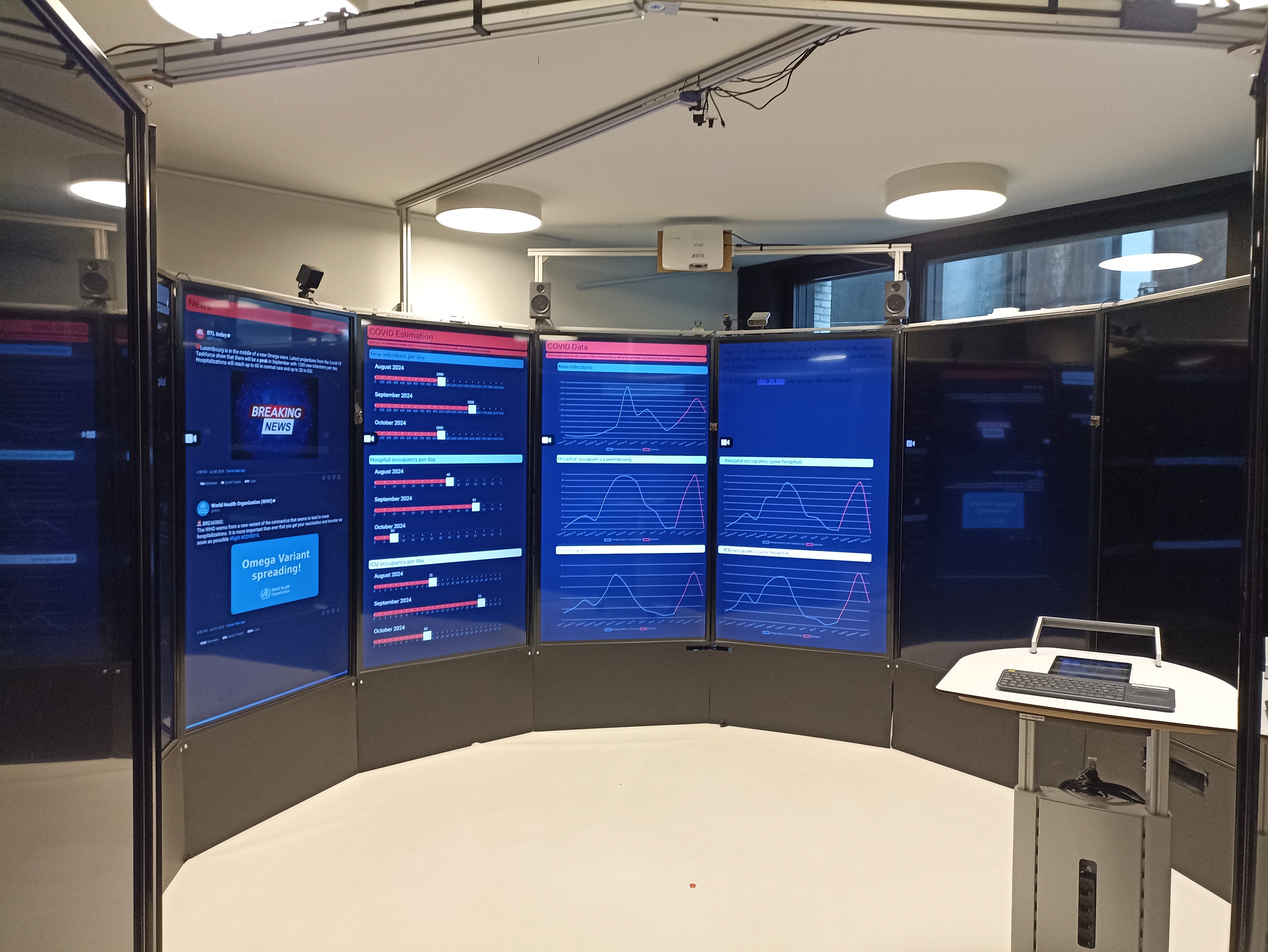}
  \caption{WSD-IA displaying the interface used for the task.} 
  \label{fig:WSD-IA}
\end{figure}

\begin{figure}[ht]
  \centering
  \includegraphics[width=0.48\textwidth]{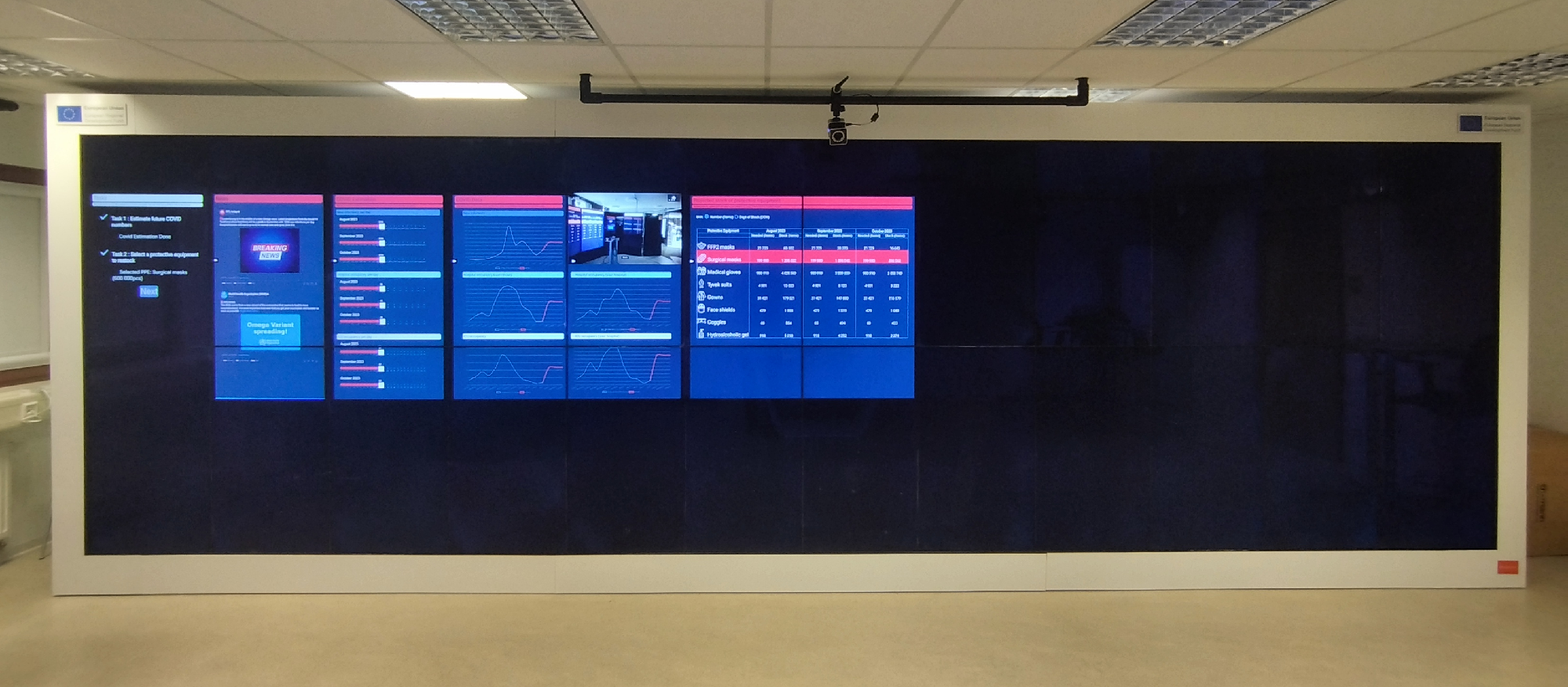}
  \caption{WSD-VW displaying a version of the interface used for the task.} 
  \label{fig:WSD-WV}

\end{figure}

%\color{black}
Two WSDs were used. First, WSD-IA (curved, diameter: $3.64$m, height: $2$m, composed of twelve 4K screens in portrait mode, among which eight are touch-enabled using infrared frames. The setup also included a height-adjustable table and a keyboard/touchpad, as shown in Figure~\ref{fig:WSD-IA}.

Then, WSD-VW (flat, width: 7m, height: 2m, total resolution $13,152\times3,872$ pixels, composed of 24 HD screens with infrared frames enabling touch (see Figure~\ref{fig:WSD-WV}). This WSD is located in a room containing three high tables (fixed-height) placed at each end of the WSD, with two mobile extended-height chairs each. The room also contains a large standard height table facing the middle of the WSD (about 3 meters away), and the display itself includes a virtual touch keyboard that appears at the bottom center of the WSD.

\subsection{Observation and analysis methods}\label{sec:obsMethods}
\begin{figure}[ht]
  \centering
  \includegraphics[width=0.48\textwidth]{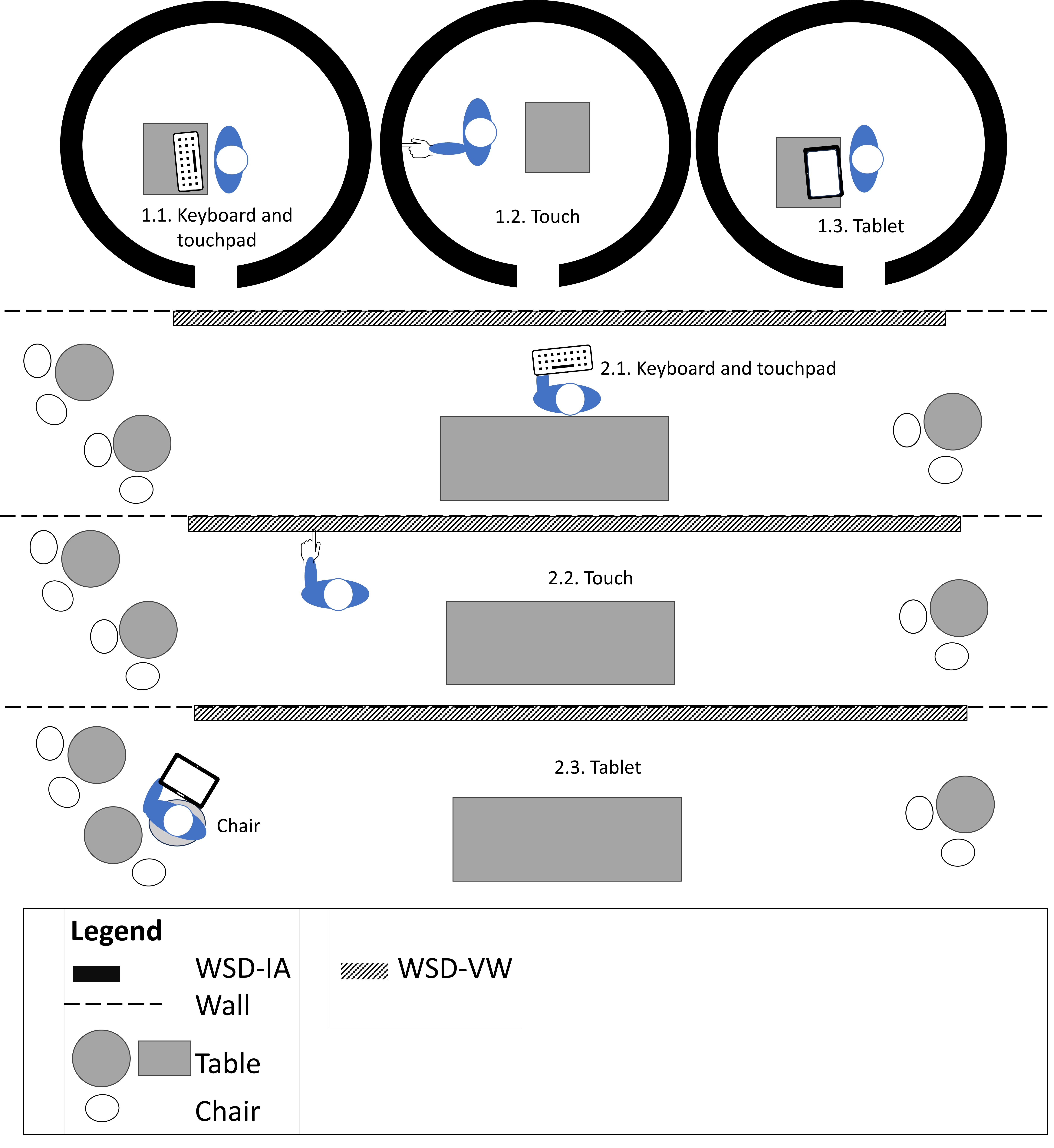}
  \caption{Experimental settings: at the top, the curved WSD-IA, at the bottom, the flat WSD-VW. The spatial relationships depicted here are illustrative and may not correspond to real-world dimensions.} 
  \label{fig:settings}
\end{figure}

Video cameras and microphones were used to record the sessions. %\color{blue}
For WSD-IA, three video cameras were used, one top camera placed in the middle of the room and attached to the roof, one in the center and attached to the top of the WSD (in front of the entrance), and one on the back, attached at the top of the entrance screens (the two screens, which can be rotated)%\color{red}
; see Figure~\ref{fig:settings} and Figure~\ref{fig:keyboard} for clarifications. \color{black} For WSD-VW, two cameras were placed in the back, in front of both ends of the display for the first user study. They were completed by a third one at the middle back for the second user study.
At the end of each session, the participant was invited to discuss and debrief with the facilitator.
We analyzed thematically the comments made by the participants, both based on the debriefings and during the sessions themselves, to identify encountered issues.
We additionally observed the moves and strategies the participants relied on.

%\color{black}

\section{Prototyping at 1:1 scale on WSD with Figma} \label{sec:XP1}

%\color{red}
During the first exploratory user study, we used Figma, a mid-fidelity web-based prototyping tool for designing, collaborating, prototyping and transmitting content~\cite{figma}. 
We chose it for its popularity~\cite{figliolia2020experiences}\cite{uidesign} and its availability as a ready-to-use web-based solution.
%\color{blue}
Figma was tested on a WSD by one designer under several experimental conditions: two WSD settings (WSD-IA, and WSD-VW) with different arrangements and surroundings, and three different interaction methods to reproduce the interface: a wireless keyboard with a touchpad, direct touch on the WSD, and a synchronized tablet.
%\color{red}

The main research question addressed by this first user study is: 
\textbf{can Figma}, as a desktop optimized tool, \textbf{be used in a WSD environment to design at 1:1 scale?}
%\color{red}

In other words, we want to identify the challenges and opportunities raised by using an existing design tool to prototype the UI of an application built for a WSD, directly on the WSD. %\color{black}

\subsection{Protocol}

The \textbf{participant} is an expert in UI design and has participated in the design of several UIs for WSDs, but had never used Figma before. She was free to stop the session whenever she wanted (e.g., when it became too difficult) or after having finished reproducing the design. Since this first user study aimed to verify the feasibility of using Figma under these conditions before carrying out more in-depth studies, a single user was deemed sufficient. Conversation guide and detailed protocol are available in supplementary material online~\cite{annex1}.

%\color{red}
Figure~\ref{fig:settings} illustrates the interaction methods tested to interact with Figma to complete the \textbf{task} as described in Section~\ref{sec:RQ}: respectively, a wireless keyboard with a touchpad, direct touch on the WSD, and a tablet synchronized with the WSD.%\color{black} 

\subsection{Results}
In general, the participant appreciated the ability to design at a 1:1 scale, regardless of the interaction method and the WSD used, with the main advantage of being able to see in real time the final rendering on the target screen.
Several issues are related to the lack of familiarity with Figma, as the use of widgets, components, and plugins seemed complicated, and were not used effectively by the participant. Also, some problems are due to or emphasized by the circular nature of WSD-IA.

\subsubsection{General observation about Figma}
The configuration of the \textbf{Figma environment} was not always suited for WSDs. For instance, the properties of a selected object are placed on the right-hand side of the display (see Figure~\ref{fig:touch}.a), the main menu is placed at the very top (see Figure~\ref{fig:touch}.b) and dialogue boxes open in the middle of the display. % (Figure~\ref{fig:FigmaIssues}.b). 
The user must also move the cursor across the entire display or physically walk to the desired location to modify, e.g., element properties (see Figure~\ref{fig:keyboard}.d), which is tiring on the long run.
Below, we discuss in more detail the issues related to each interaction method.

\subsubsection{Interacting with a wireless keyboard with a touchpad}
The session lasted one hour for WSD-IA and ten minutes for WSD-VW. 
On both WSDs, the participant would sometimes have a hard time finding the cursor on the large display.

\begin{figure}[htbp]
  \centerline{\includegraphics[scale=0.52]{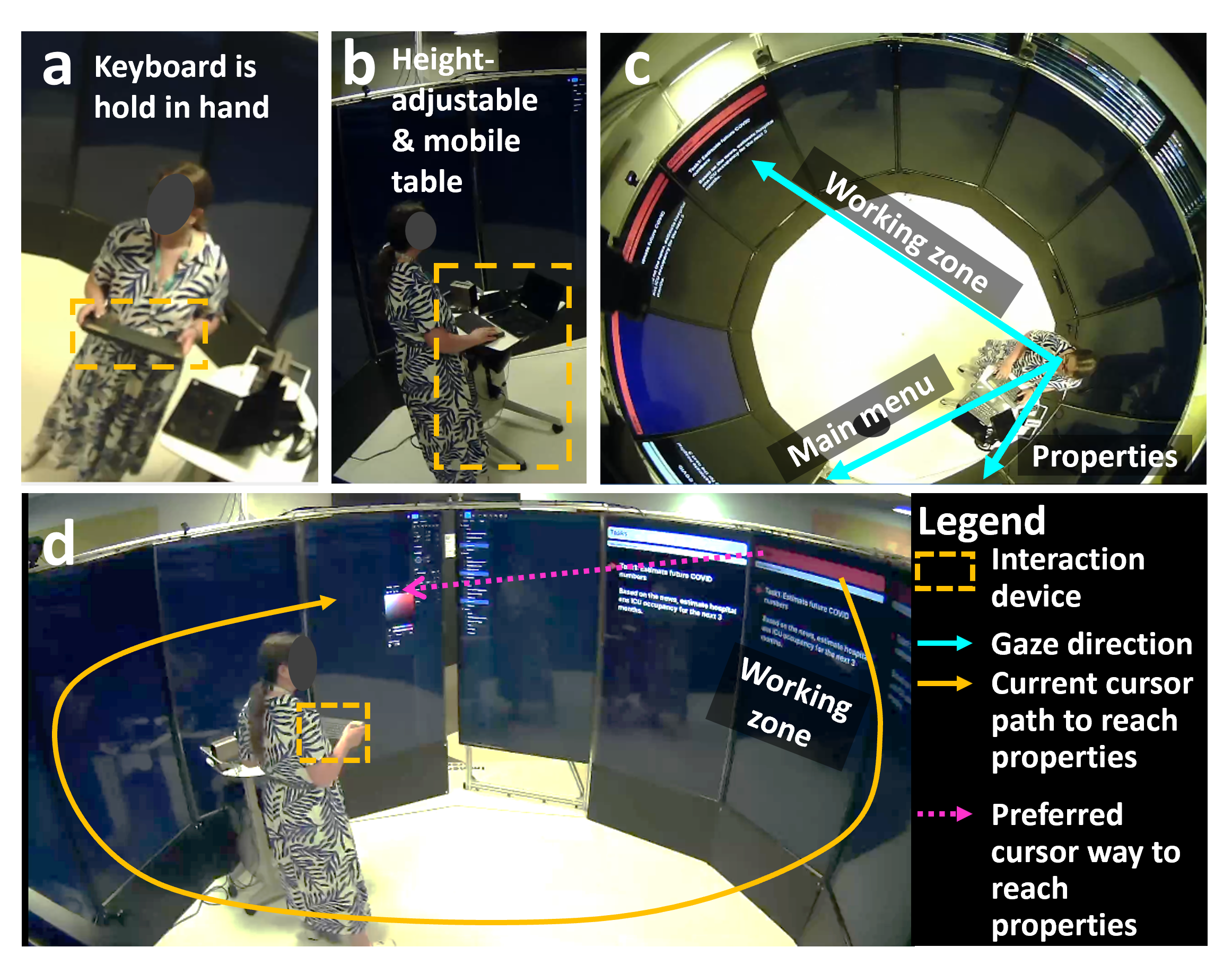}}
  \caption{Observations made when interacting with Figma and keyboard+touchpad. a)~At first, the participant held the keyboard. b)~Use of a table to put down the keyboard. c) Lots of head rotations to see all the important areas. d)~WSD-IA does not allow you to cross directly from the left screen to the right screen.} 
  \label{fig:keyboard}
\end{figure}

Concerning the \textit{WSD-IA}, the menu and items list were displayed on the leftmost screen (one of the two screens also used as a door), and the selected item's properties on the rightmost screen (the other screen forming the side by side door). To avoid turning her head from the extreme left to the extreme right too often, the participant closed the doors to look at them both at the same time. %, see Figure~\ref{fig:FigmaIssues}.a. 
She also placed herself to create an angle that allowed her to see the menu, the properties of the selected object, and the work area at a glance (see Figure~\ref{fig:keyboard}.c). 
As the session was short, and all UI elements were tightly grouped on the left, the position was acceptable. But the user could not maintain this position while working in the middle. 
In this configuration (menu on the left and properties on the right), the participant turned her head and body a lot, which could possibly be painful and exhausting.
At first, the participant carried the interaction device, see Figure~\ref{fig:keyboard}.a. After 15 minutes, she felt tired and placed it on a table, see Figure~\ref{fig:keyboard}.b.
Another problem was the impossibility to move the cursor directly from the WSD's leftmost side to its rightmost side. The participant had to move the cursor all the way around the WSD, which is tiring, despite the physical proximity of both sides of the WSD due to its circular arrangement (see Figure~\ref{fig:keyboard}.d). To avoid turning her head too much, the participant did not follow the cursor with her eyes when it was behind her back.

In the \textit{WSD-VW} condition, the font size of the Figma interface was problematic. The flat WSD-VW was too wide to read text labels when standing at the opposite side of the display. Hence, to modify a property's value, the user had to walk frequently across the WSD-VW to the properties area, where she rested on a table next to it. %, see Figure~\ref{fig:FigmaIssues}.e.
Then she leaned on the middle table for comfort and stayed at a certain distance from WSD-VW to get an overview. 
The fatigue due to walking around, eye strain due to the text size, and carrying the keyboard led the user to stop the test after ten minutes. Because the WSD-IA is twice as large as the WSD-VW in terms of horizontal resolution, 
the virtual navigation felt more painful in WSD-IA. This is also the case because the cursor and its progress were always visible in WSD-VW, and the session was shorter.

\subsubsection{Interacting using direct touch on the WSD}
The session with WSD-IA lasted twenty minutes, and the session with WSD-VW was interrupted after ten minutes.

\begin{figure}[htbp]
    \centerline{\includegraphics[scale=0.39]{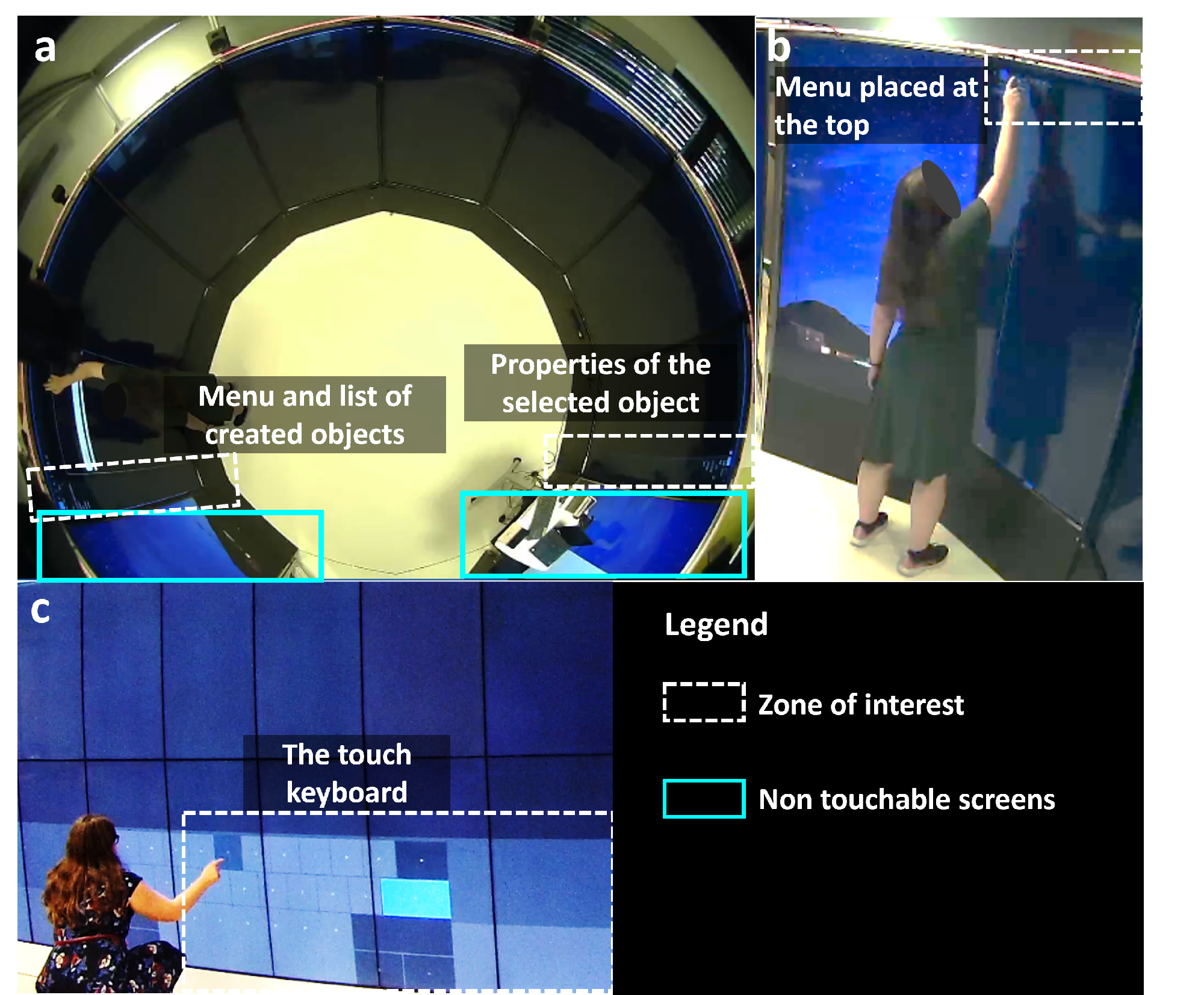}}
    \caption{Observations done when interacting with Figma and touch. a)~The touch space is occupied by the Figma interface on the left (list of created objects) and the right (properties). b)~The menu is too high. c)~WSD-VW's touch keyboard is not comfortable; the user had to crouch.} 
    \label{fig:touch}
\end{figure}

In the \textit{WSD-IA} condition, to manage physical fatigue (neck strain and gorilla arm syndrome~\cite{hansberger2017dispelling}), the user tried to work at a lower scale by zooming in on the work area without minimizing the Figma window. Although the menu remained too high and objects' properties too far away, objects could be moved with smaller movements and 
were better placed relative to the user's field of view, generating less neck pain. Hence, the advantage of working at 1:1 scale was temporarily lost.
Only the middle eight screens of WSD-IA support touch, so we resized the browser window used for Figma to fit within these screens. 
As Figma's interface elements (list of objects and properties) take up space themselves at both ends of the window, the design space was further reduced, by about one extra screen for interface elements from both ends combined. Hence, the total design space available to the participant was reduced compared to the actual interface they had to reproduce. Consequently, the alignment with the display tiles could not be maintained, as shown in Figure~\ref{fig:touch}.a.
When using touch, the participant had trouble moving an object across tiles.
The wireless keyboard was used to input text or values and was either held by the participant or placed on the table.

On \textit{WSD-VW}, the properties panel was too far away from the work area, but unlike WSD-IA, when a property was changed on WSD-VW, the result was not visible from the user's position. 
So, she stepped back to check, e.g., whether the font size is large enough.
The top menu was out of reach, and WSD-VW's virtual keyboard was not suitable for entering more than one word due to its design (position at the bottom and large size, see Figure~\ref{fig:touch}.c). After ten minutes of use, the participant complained from the gorilla arm syndrome.

\subsubsection{Interacting on a synchronized tablet}
The same Figma project was loaded onto the tablet and onto the WSD. The UI elements were created, moved and adjusted on the tablet. 

\begin{figure}[htbp]
    \centerline{\includegraphics[scale=0.48]{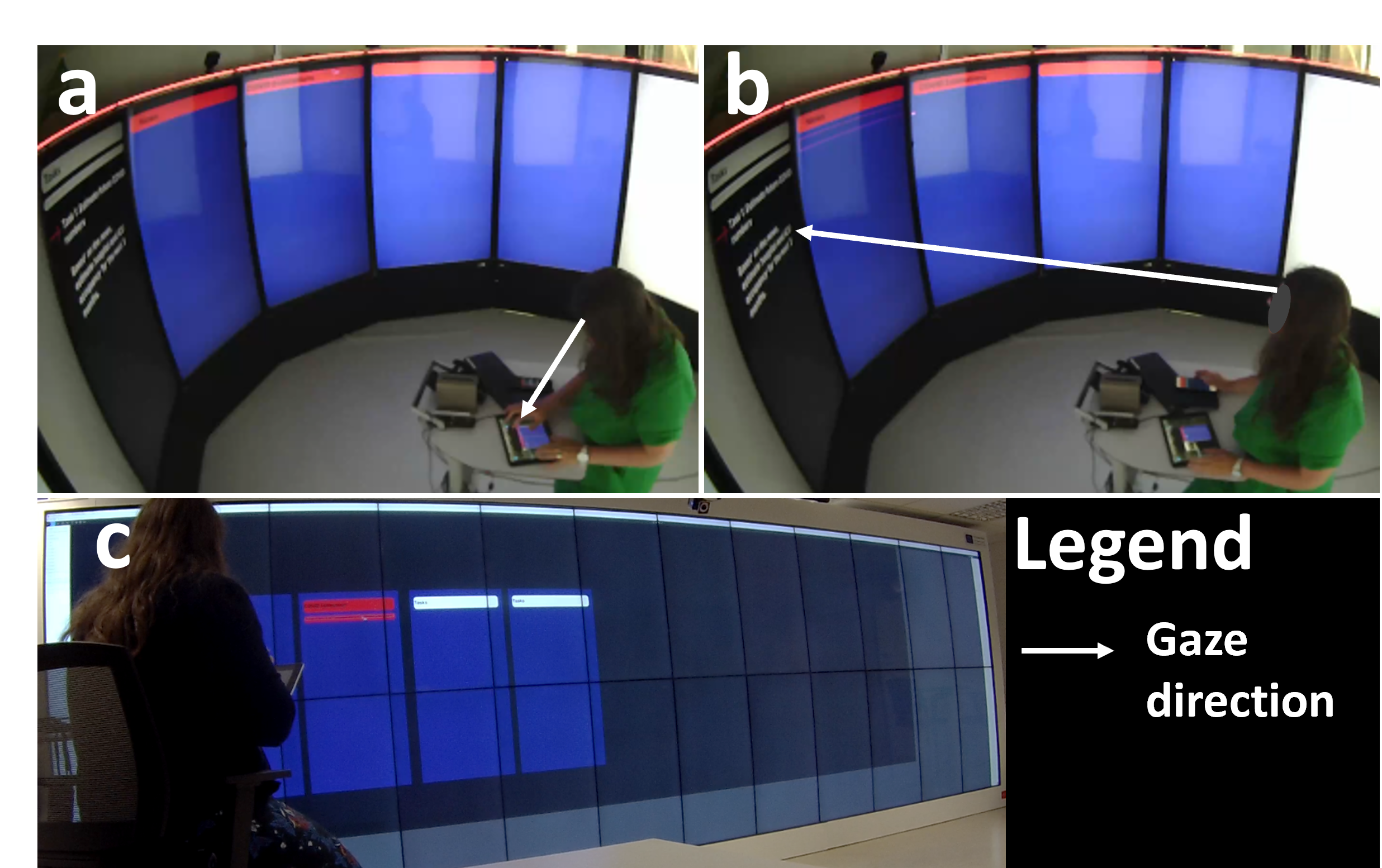}}
    \caption{Observations done when interacting with Figma and a tablet. a)~The participant modified the prototype on the tablet. b)~Then, the participant checked the result on the WSD. c)~The participant sat on a chair.} 
    \label{fig:tablet}
\end{figure}

We observed that the participant mainly looked at the tablet to add UI elements, move them around and set parameters, see Figure~\ref{fig:tablet}.a. Then, the participant looked at the WSD to check, e.g., the position and size of the UI elements, the readability of text, and colors, see Figure~\ref{fig:tablet}.b. 
A main issue was the impossibility to select several UI elements at the same time on the tablet, as they are superimposed. 
The session with WSD-IA lasted ninety minutes, whereas the session with WSD-VW was interrupted after twenty minutes.

On \textit{WSD-IA}, the user had difficulties to position the UI prototype on the WSD correctly. Although the additions and changes to UI elements were synchronized between the tablet and the WSD, the viewports were independent, so positioning the UI had to be done from the WSD directly. %, as the position on the WSD was not synchronized with the tablet, although the modification of UI elements was kept in sync. That means that the WSD and the tablet show two independent viewports of the same design. 
This led to the use of an extra wireless keyboard equipped with a touchpad. 
This happened when the participant closed the project and reopened it. The designed UI was centered horizontally and vertically on the WSD. So, the participant wanted to properly reposition the designed UI to continue the design.
The participant also used the touchpad to select a group of UI elements to save them as a new reusable UI element. She placed the tablet on the height-adjustable and mobile table. She felt that WSD-IA and tablet configuration was the most comfortable.

On \textit{WSD-VW}, the participant sat down and placed the tablet on the table, see Figure~\ref{fig:tablet}.c.
But, as the table was not well positioned and too heavy to be moved, she preferred to hold the tablet in her hand, which was tiresome. 

\subsection{Preliminary observations}\label{sec:XP1obs}
The duration of the test sessions varied widely, from ten to 90 minutes.
The most comfortable condition seems to be WSD-IA with a tablet and a height-adjustable and mobile table. But the problem of multiple selection and correct positioning of the prototype on the WSD needs to be solved. 
Overall, the main issues were: 
\begin{enumerate*}[label=(I\arabic*)]
    \item physical fatigue,
    \item reachability of Figma elements,
    \item readability of the Figma interface, 
    \item the vast interaction surface,
    \item when a project is reopened, objects are moved to the middle, and 
    \item the partial occlusion of the WSD by Figma's UI elements, which is not a perfect 1:1 scale. 
\end{enumerate*}

To tackle these issues, we outline several design solutions.
(I1)~could be reduced by managing the physical environment and providing a height-adjustable and movable table to place the interacting device, chairs, and by supporting interaction at a distance.
(I2)~could be improved by offering floating context menus and value input fields, by opening dialogue boxes close to the work area or by using a smaller interaction device as a tablet or a laptop.
For (I3) and (I4), the size of the design tool interface elements should be adapted. 
For (I4), a bigger cursor should be used as well as pointing facilitation techniques~\cite{BALAKRISHNAN2004857} to reach the opposite end of the WSD.
(I5)~could be solved by fixing the created objects in their positions and reloading them in exactly the same position. 
To achieve 1:1 scale (I6), the design tool interface should be hidable or movable.

\begin{sidewaystable*}
\caption{%\color{red}
Main issues and possible solutions.%\color{black}
}
\centering
\resizebox{\textwidth}{!}{%
\begin{tabular}{@{}lllllllll@{}}
\hline
\multirow{2}{*}{} &
  \multirow{2}{*}{Issues} &
  \multicolumn{2}{c}{\textbf{Keyboard/touchpad}} &
  \multicolumn{2}{c}{\textbf{Touch}} &
  \multicolumn{2}{c}{\textbf{Tablet}} &
  \multirow{2}{*}{Possible solutions} \\ 
 &
   &
  WSD-IA &
  WSD-VW &
  WSD-IA &
  WSD-VW &
  WSD-IA &
  WSD-VW &
   \\ 
   \hline
\multirow{2}{*}{I1} &
  \multirow{2}{*}{Physical fatigue} &
  Yes &
  Yes &
  Yes &
  Yes &
  No &
  Yes &
  \multirow{2}{*}{\begin{tabular}[c]{@{}l@{}}A height-adjustable movable table on \\ 
                                            which to place the interaction device, a chair \\ 
                                            for sessions longer than 90 minutes.\\
                                            Interacting at a distance.
                                            \end{tabular}} \\
 &
   &
  \begin{tabular}[c]{@{}l@{}}frequent head and \\ torso rotations \\ \end{tabular} &
  \begin{tabular}[c]{@{}l@{}}hand-held, \\ much walking\end{tabular} &
  \begin{tabular}[c]{@{}l@{}}Menu too high,\\ much walking\end{tabular} &
  \begin{tabular}[c]{@{}l@{}}Menu too high,\\ much walking, touch \\ keyboard too low\end{tabular} &
  placed on a table &
  hand-held &
   \\
   \hline
\multirow{2}{*}{I2} &
  \multirow{2}{*}{\begin{tabular}[c]{@{}l@{}}Figma's elements \\ reachability\end{tabular}} &
  Yes &
  Yes &
  Yes &
  Yes &
  No &
  No &
  \multirow{2}{*}{\begin{tabular}[c]{@{}l@{}}Floating context menus and value input fields\\ 
                                            near the working zone, allow the participant\\
                                            to move across the display without scrolling, \\
                                            open dialogue boxes near the working zone,\\
                                            and use a smaller interaction device.\\
                                            \end{tabular}} \\[31pt]
 &
   &
  \begin{tabular}[c]{@{}l@{}}Properties too \\ far from the \\ working zone\end{tabular} &
  \begin{tabular}[c]{@{}l@{}}Properties too \\ far from the \\ working zone\end{tabular} &
  \begin{tabular}[c]{@{}l@{}}Properties too \\ far from the \\ working zone\end{tabular} &
  \begin{tabular}[c]{@{}l@{}}Properties too \\ far from the \\ working zone\end{tabular} &
   &
   &
   \\
   %\midrule
   \hline
\multirow{2}{*}{I3} &
  \multirow{2}{*}{Readability} &
  No &
  Yes &
  No &
  Yes &
  No &
  No &
  \multirow{2}{*}{\begin{tabular}[c]{@{}l@{}}Adapt the size of the Figma elements, \\
                                            use a curved WSD instead.\end{tabular}} \\[8pt]
 &
   &
   &
  \begin{tabular}[c]{@{}l@{}}UI element not \\ visible when\\ modifying its \\ parameters.\\ Figma interface \\ not readable.\end{tabular} &
   &
  \begin{tabular}[c]{@{}l@{}}UI element not \\ visible when\\ modifying its\\ parameters\end{tabular} &
   &
   &
   \\
   \hline
\multirow{2}{*}{I4} &
  \multirow{2}{*}{\begin{tabular}[c]{@{}l@{}}Huge interaction \\ surface\end{tabular}} &
  Yes &
  Yes &
  No &
  No &
  No &
  No &
  \multirow{2}{*}{\begin{tabular}[c]{@{}l@{}}Adapt the size of the Figma elements, \\
                                            use a curved WSD instead.\\  
                                            Bigger cursor, pointer acceleration to \\
                                            reach the opposite side of the display.\end{tabular}} \\ 
 &
   &
  \begin{tabular}[c]{@{}l@{}}It takes a long \\ time to reach \\ the extremities.\end{tabular} &
  \begin{tabular}[c]{@{}l@{}}It takes a long\\ time to reach\\ the extremities.\end{tabular} &
   &
   &
   &
   &
   \\
   %\midrule
   \hline
\multirow{2}{*}{I5} &
  \multirow{2}{*}{\begin{tabular}[c]{@{}l@{}}Not fixed position \\ of UI elements\end{tabular}} &
  Yes &
  Yes &
  Yes &
  Yes &
  Yes &
  Yes &
  \multirow{2}{*}{\begin{tabular}[c]{@{}l@{}}Fix the UI elements in their position and \\
                                            restore their position after reloading \\
                                            the project. The Figma design space should\\
                                            match the targeted WSD.
                                            \end{tabular}} \\ [21pt]
 &
   &
  \multicolumn{6}{c}{When reopening the project.} &
   \\
   \hline
\multirow{2}{*}{I6} &
  \multirow{2}{*}{1:1 scale not perfect} &
  No &
  No &
  Yes &
  No &
  No &
  No &
  \multirow{2}{*}{\begin{tabular}[c]{@{}l@{}}Possibility to conceal or move \\
                                            the Figma interface elements\\
                                            such as the menus.\end{tabular}} \\ 
 &
   &
   &
   &
  \begin{tabular}[c]{@{}l@{}}Part of the necessary \\ design surface masked\\ by Figma elements.\end{tabular} &
   &
   &
   &
   \\ 
   \hline
\end{tabular}%
}
\label{tab:IssuesAndSolutions}
\end{sidewaystable*}

Overall, issues (I2), (I3), (I5) and (I6) show that Figma may not be adapted to prototype at 1:1 scale on WSDs, and that another tool incorporating the outlined design solutions, might be better suited.

%\color{red}
\section{Comparative analysis of existing design tools}\label{sec:benchmark}%Features matrix}
This first user study enabled us to identify the problems with using Figma as a 1:1 scale design tool for WSDs. Based on these results, we then set up a list of required features for such a tool. Below, we first list these features and then analyze 19 different design tools, meant for UI design for computers and smartphones, taking into account the listed features. 

Based on our observations, we identified the following main features of tools suited for designing at 1:1 scale on WSD:
\begin{itemize}
    \item The main menu must be reachable from the work area (not at the top or bottom) [related to I1, I2, I4].
    \item The properties of the selected object must be reachable from the work area [related to I1, I2, I4].
    \item Pop-ups must open near the work area [related to I1, I2, I4].
    \item The font size of the design tool UI elements must be large enough (or adjustable) [related to I3].
    \item It must be possible to conceal or move the design tool interface elements, i.e., menu, list of created objects, and properties of the selected object [related to I6].
    \item It must be possible to interact from a synchronized tablet [needed for our user study].
    \item When reopening a previous project, it must preserve the positions of the created interface elements, as they were previously. [related to I5]
    \item The workspace size must be large or adaptable to the typically high resolution of WSDs.
\end{itemize}

Based on this features list, we analyzed 19 design tools. 
We performed a search using the engine Google with the keywords ``design tool'', ``mockup tool'' and ``UI design tool'' between July and December 2024. We collected the design tools from direct results or from online articles about such tools~\cite{uidesign}\cite{uxtools}. We then eliminated the tools that did not support the creation of mock-ups, and the ones that could not be tested and used for free. In the end, we selected 19 tools, which were analyzed against the listed features. Many tools set fixed positions for their menus, and for object properties panels. None of the tools in the selection enabled users to conceal or move the design tool's interface elements.

As our WSDs run on a Linux operating system, we also had to narrow down our list to multiplatform tools, which indirectly led us to mainly consider web-based solutions.

Table~\ref{tab:benchmark} lists the design tools and our analysis of their characteristics against the desirable features. 
Concerning the maximum surface size of the workspace, this information was rather complex and at times impossible to find for some design tools, so some figures listed in the table might not be accurate.
As most of the examined tools provide a web-based version, thus giving access to the zoom functionality of the web browser, we did not check the font size of the design tool's interface.

\begin{sidewaystable*}
\caption{%\color{red}
Features matrix of the UI design tools. Cells in gray highlight instances where features or requirements we identified are supported by the corresponding design tool. NA: not applicable. * No source for inVision as the design tool was stopped after our review. Adobe XD is now in maintenance mode. ** E.g., delete. %\color{black}
}
\small
\resizebox{.9\linewidth}{!}{%
\renewcommand{\arraystretch}{0.7}
\begin{tabular*}{\textwidth}{llllllllll}%{\textwidth}
\textbf{Name}            
& \begin{tabular}[c]{@{}l@{}}\textbf{Max surface} \\ \textbf{size}\end{tabular}    
& \textbf{Supported platforms}
& \begin{tabular}[c]{@{}l@{}}\textbf{Main menu} \\ \textbf{position}\end{tabular}         
& \begin{tabular}[c]{@{}l@{}}\textbf{Object properties} \\ \textbf{position}\end{tabular}         
& \begin{tabular}[c]{@{}l@{}}\textbf{Dialog} \\ \textbf{boxes} \\ \textbf{position}\end{tabular} 
& \begin{tabular}[c]{@{}l@{}}\textbf{Reopens at the} \\\textbf{same position}\end{tabular} 
& \begin{tabular}[c]{@{}l@{}}\textbf{Retractable} \\ \textbf{interface}\end{tabular} 
& \begin{tabular}[c]{@{}l@{}}\textbf{Synchronization} \\ \textbf{with a tablet}\end{tabular}        
& \textbf{Selected}                       \\
\hline
Adobe XD*~\cite{Adobe_2024}         
    & 50,000x50,000 px                                               
    & \cellcolor[HTML]{C0C0C0}\begin{tabular}[c]{@{}l@{}}macOS, Windows,\\ Android, iOS, \\ browsers (Chrome, \\ Firefox, Edge, Safari)\end{tabular} 
    & \begin{tabular}[c]{@{}l@{}}fixed on the top \\ and left\end{tabular}  
    & fixed on the right                                                & middle                                                            & unknown  
    & unknown
    & unknown                                                           & no                       \\
\hline
Axure RP~\cite{Axure}        
    & 20,000 px                                                      
    & \begin{tabular}[c]{@{}l@{}}Windows, macOS, \\ Android, iOS\end{tabular}                                                    & fixed on the top                                                  
    & fixed on the right                                                & middle                                                            & unknown                                                           & unknown                                                           & unknown                                                          
    & no                       \\
\hline
Balsamiq~\cite{Balsamiq}       
    & 20,000x20,000 px                                               
    & \cellcolor[HTML]{C0C0C0}\begin{tabular}[c]{@{}l@{}}Windows, macOS, \\ browsers\end{tabular}                                     
    & fixed on the top                                                  
    & fixed on the right                                                
    & middle                                                            
    & \cellcolor[HTML]{C0C0C0}yes   
    & \cellcolor[HTML]{C0C0C0}\begin{tabular}[c]{@{}l@{}}partially +\\ preview mode \end{tabular}   
    & \cellcolor[HTML]{C0C0C0}\begin{tabular}[c]{@{}l@{}}possible with the \\ cloud version%, \\ not tested for \\ installed versions
        \end{tabular}               
    & no                       \\
\hline
Bubble~\cite{Bubble}          
    & max width 5,000 px                                             
    & \cellcolor[HTML]{C0C0C0}browsers                                  
    & fixed on the left                                                 
    & \cellcolor[HTML]{C0C0C0}floating and movable                      
    & unknown                                                           
    & no                   
    & \cellcolor[HTML]{C0C0C0}\begin{tabular}[c]{@{}l@{}}partially +\\ preview mode \end{tabular}   
    & \cellcolor[HTML]{C0C0C0}\begin{tabular}[c]{@{}l@{}}Possible with the \\ opening of the \\ same project, \\ but lag\end{tabular}      
    & \cellcolor[HTML]{9B9B9B}yes \\
\hline
Canva~\cite{Canva}          
    & 8000 x 3125 px                                                 
    & \cellcolor[HTML]{C0C0C0}\begin{tabular}[c]{@{}l@{}}browsers, Windows, \\ macOS, iOS, \\ Android\end{tabular}    
    & fixed on the left                                 
    & \cellcolor[HTML]{C0C0C0}\begin{tabular}[c]{@{}l@{}}fixed on the top\\ some actions \\ attached to the \\ selected object**)\end{tabular}   
    & middle                                                            
    & no  
    & \cellcolor[HTML]{C0C0C0}\begin{tabular}[c]{@{}l@{}}partially +\\ preview mode \end{tabular}   
    & \cellcolor[HTML]{C0C0C0}\begin{tabular}[c]{@{}l@{}}Possible with the \\ cloud version
    \end{tabular}               
    & no                       \\
\hline
Excalidraw~\cite{Excalidraw}     
    & unknown                                                           
    & \cellcolor[HTML]{C0C0C0}browsers                                  
    & fixed on the top                                                  
    & fixed on the left                                                 
    & unknown                                                           
    & \cellcolor[HTML]{C0C0C0}yes         
    & \cellcolor[HTML]{C0C0C0}partially   
    & \cellcolor[HTML]{C0C0C0}\begin{tabular}[c]{@{}l@{}}using collaboration \\ feature\end{tabular}                              
    & no                       \\
\hline
Figma~\cite{figma}          
    & no limit                                                       
    & \cellcolor[HTML]{C0C0C0}browsers                                 
    & fixed on the top                                                  
    & fixed on the right 
    & middle                                                            
    & no  
    & \cellcolor[HTML]{C0C0C0}\begin{tabular}[c]{@{}l@{}}partially +\\ preview mode \end{tabular}   
    & \cellcolor[HTML]{C0C0C0}\begin{tabular}[c]{@{}l@{}}Possible with the \\ opening of the\\ same project or \\ using collaboration \\ feature\end{tabular}   
    & no                       \\
\hline
Framer~\cite{Framer}         
    & max width 1280px                                               
    & \cellcolor[HTML]{C0C0C0}\begin{tabular}[c]{@{}l@{}}browsers, macOS, \\ Windows\end{tabular}                                      
    & \begin{tabular}[c]{@{}l@{}}fixed on the top \\ and left\end{tabular}  & fixed on the right                            
    & middle                                                            
    & no          
    & \cellcolor[HTML]{C0C0C0}\begin{tabular}[c]{@{}l@{}}preview mode \end{tabular}   
    & \cellcolor[HTML]{C0C0C0}\begin{tabular}[c]{@{}l@{}}Possible with the \\ cloud version, %\\ not tested for \\ installed versions
\end{tabular}               
    & no                       \\
\hline
InVision*        
    & unknown                                                           
    & \cellcolor[HTML]{C0C0C0}browsers                                  
    & \begin{tabular}[c]{@{}l@{}}fixed on the \\ bottom middle\end{tabular} 
    & \cellcolor[HTML]{C0C0C0}\begin{tabular}[c]{@{}l@{}}attached to the \\ selected object\end{tabular}                                
    & top left                                                          
    & no
    & unknown   
    & \cellcolor[HTML]{C0C0C0}\begin{tabular}[c]{@{}l@{}}Possible with the\\ opening of the \\ same project,\end{tabular}                  
    & \cellcolor[HTML]{9B9B9B}yes \\
\hline
JustInMind~\cite{JustinMind}      
    & unknown                                                           
    & Windows, macOS                                                    
    & fixed on the top                                                  
    & fixed on the right                                                
    & middle                                                            
    & unknown 
    & unknown  
    & \cellcolor[HTML]{C0C0C0}\begin{tabular}[c]{@{}l@{}}using collaboration \\ feature\end{tabular}                               
    & no                       \\
\hline
Miro~\cite{Miro}            
    & unknown                                                             & \cellcolor[HTML]{C0C0C0}\begin{tabular}[c]{@{}l@{}}browsers, iOS, \\ Android, macOS, \\ Windows\end{tabular}                     
    & fixed on the left                                                 
    & \cellcolor[HTML]{C0C0C0}\begin{tabular}[c]{@{}l@{}}attached to the \\ selected object\end{tabular}                                 
    & middle                                                            & \cellcolor[HTML]{C0C0C0}yes 
    & \cellcolor[HTML]{C0C0C0}\begin{tabular}[c]{@{}l@{}}preview mode \end{tabular}   
    & \cellcolor[HTML]{C0C0C0}\begin{tabular}[c]{@{}l@{}}Possible with the \\ opening of the \\ same project, \\ or using \\ collaboration feature\end{tabular} 
    & \cellcolor[HTML]{9B9B9B}yes \\
\hline
MockFlow~\cite{MockFlow}        
    & unknown                                                           
    & \cellcolor[HTML]{C0C0C0}\begin{tabular}[c]{@{}l@{}}browser, macOS, \\ Windows\end{tabular}                                      
    & fixed on the left                                                 & fixed on the left                                                 
    & unknown                                                           
    & no        
    & no   
    & unknown                                                           
    & no                       \\
\hline
Mockplus~\cite{Mockplus}        
    & no limit                                                       
    & \cellcolor[HTML]{C0C0C0}\begin{tabular}[c]{@{}l@{}}browsers, Android, \\ iOS, Windows, \\ macOS\end{tabular}                     
    & \begin{tabular}[c]{@{}l@{}}fixed on the top\\ and left\end{tabular}
    & fixed on the right                                                
    & middle                                                            
    & \cellcolor[HTML]{C0C0C0}yes    
    & \cellcolor[HTML]{C0C0C0}\begin{tabular}[c]{@{}l@{}}preview mode \end{tabular}   
    & \cellcolor[HTML]{C0C0C0}\begin{tabular}[c]{@{}l@{}}using collaboration \\ feature (only \\ visualization, no \\ modification)\end{tabular}
    & no                       \\
\hline
Penpot~\cite{penpot}          
    & no limit                                                       
    & \cellcolor[HTML]{C0C0C0}browsers                                  
    & fixed on the top                                                  
    & fixed on the right                                                
    & middle                                                            
    & \cellcolor[HTML]{C0C0C0}yes    
    & \cellcolor[HTML]{C0C0C0}\begin{tabular}[c]{@{}l@{}}yes +\\ preview mode \end{tabular}   
    & \cellcolor[HTML]{C0C0C0}\begin{tabular}[c]{@{}l@{}}using collaboration \\ feature\end{tabular}                               
    & no                       \\
\hline
Proto.io~\cite{protoio}       
    & unknown                                                           
    & \cellcolor[HTML]{C0C0C0}\begin{tabular}[c]{@{}l@{}}browsers (Chrome, \\ Safari, Firefox)\end{tabular}                           
    & \begin{tabular}[c]{@{}l@{}}fixed on the \\ right\end{tabular}     
    & \cellcolor[HTML]{C0C0C0}\begin{tabular}[c]{@{}l@{}}fixed on the right\\ some actions \\ attached to the \\ selected object\end{tabular} 
    & middle                                                            
    & no     
    & \cellcolor[HTML]{C0C0C0}\begin{tabular}[c]{@{}l@{}}preview mode \end{tabular}   
    & unknown                                                           
    & no                       \\
\hline
ProtoPie~\cite{protopie} 
    & unknown                                                           
    & Windows                                                           
    & fixed on the top                                                  
    & fixed on the right                                                
    & middle                                                            
    & unknown  
    & unknown   
    & unknown                                                           
    & no                       \\
\hline
Sketch~\cite{Sketch}          
    & unknown                                                           
    & macOS                                                             
    & fixed on the top                                                  
    & fixed on the right                                                
    & unknown                                                           
    & unknown
    & unknown
    & unknown                                                           
    & no                       \\
\hline
UXPin~\cite{UXPin}           
    & 25000x25000 px                                                 
    & \cellcolor[HTML]{C0C0C0}\begin{tabular}[c]{@{}l@{}}browsers, macOS, \\ Windows\end{tabular}                                      
    & fixed on the left                                                 
    & fixed on the right                                                
    & middle                                                            
    & \cellcolor[HTML]{C0C0C0}yes  
    & \cellcolor[HTML]{C0C0C0}\begin{tabular}[c]{@{}l@{}}preview mode \end{tabular}   
    & \cellcolor[HTML]{C0C0C0}\begin{tabular}[c]{@{}l@{}}Possible with the \\ opening of the \\ same project, \\ but lag\end{tabular}      
    & no                       \\
\hline
Webflow~\cite{Webflow}         
    & \begin{tabular}[c]{@{}l@{}}max width \\ 10,000 px\end{tabular} 
    & \cellcolor[HTML]{C0C0C0}browsers                  
    & fixed on the left                                                 
    & fixed on the right                                                
    & middle                                                            
    & no       
    & \cellcolor[HTML]{C0C0C0}\begin{tabular}[c]{@{}l@{}}preview mode \end{tabular}   
    & unknown                                                           
    & no                      \\
   \hline
\end{tabular*}%
}
\label{tab:benchmark}
\end{sidewaystable*}

Based on the features they offer, we pre-selected three multi-platform design tools (Bubble, inVision, and Miro) to test them more in-depth on one WSD (WSD-IA), with the three interaction methods used in the user study, to ensure their usability for the second user study.
With inVision, we ran into issues when using touch on our WSD, so we had to reject it. Additionally, we noticed that the solution was discontinued shortly after our tests.
As for Bubble, we experienced some synchronization latency when using the tablet, which is why we discarded it as well.
We therefore chose Miro as design tool, because it satisfied most of the requirements, including running in a web browser, showing object properties in a context menu, reopening a project at the same position, retractable interface by using the preview mode, and offering the option to open the same project on a tablet as on the WSD.

\section{Prototyping at 1:1 scale on WSD with Miro}\label{sec:XP2}
Based on the findings from the first user study and the feature matrix, we ran a second user study with some modifications in the protocol compared to the first study, as explained below.

The research questions are adapted from the first study. RQ1:~\textbf{Can an existing design tool}, that was conceived for desktops, \textbf{be used in a WSD environment to design at 1:1 scale?} And RQ2: \textbf{What are the main guidelines to support design at 1:1 scale in a WSD environment?}

\subsection{Protocol}
%To better suit the above features for the WSD environment, we made some adaptations. 
Miro was tested as a mockup tool for WSDs by two designers under several conditions: two WSD settings (WSD-IA, and WSD-VW), and three different interaction methods: (1) a wireless keyboard with a touchpad, (2) touch on the WSD completed by a wireless keyboard for entering textual and numeric values, and (3) a synchronized tablet.

Sessions were limited to one hour. But, participants were free to stop the session earlier, if they felt too tired, or uncomfortable, or if they completed the task.

The two \textbf{participants} were respectively an expert in UI design (p2) and in visualization design (p1). They both participated in the design of several UIs for WSDs, although (p2) is more experienced in UI design per se. They both had already used Miro beforehand as a collaborative mind mapping tool, but not for UI design, with p2 who was competent using Miro and p1 self-declaring as more novice. 

Like in the first study, the \textbf{system} used for this second user study consists of a touch-enabled WSD displaying the design tool, Miro in this case, through the Chrome web browser.
%%in full-screen mode
The zoom level in the browser was set to 200\%, for the main menu items (located on the left and vertically centered) to be wide enough to be usable.
We used the same two WSDs as previously: WSD-IA (curved) and WSD-VW (flat). 

\begin{figure}
    \centering
    \includegraphics[width=0.95\linewidth]{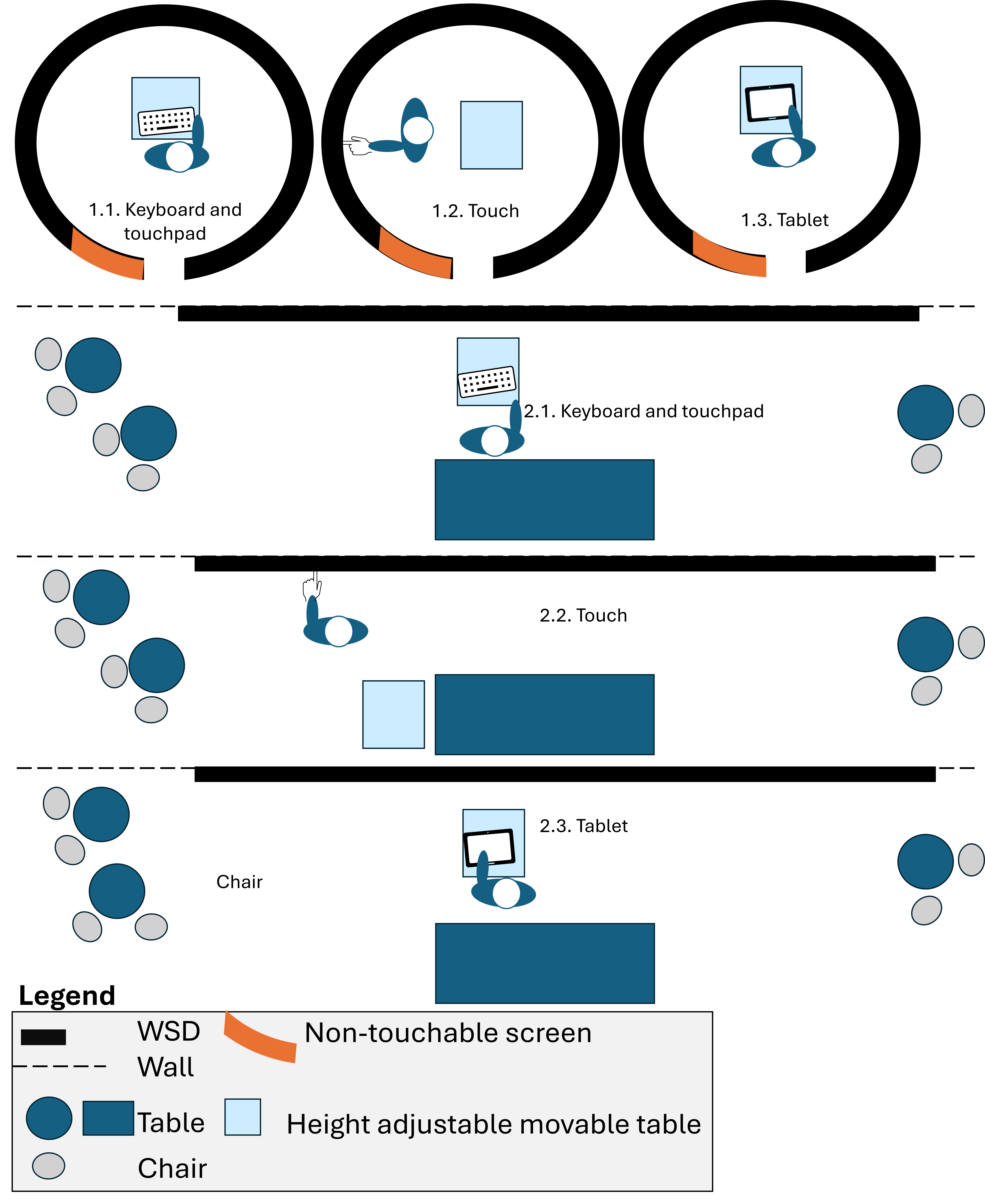}
    \caption{Experimental settings for the second study: on the top the curved WSD-IA, on the bottom the flat WSD-VW. The spatial relationships depicted in this figure are illustrative and may not correspond to real-world dimensions.}
    \label{fig:system2}
\end{figure}

Learning from the first user study, some adaptations were made on both WSDs. 
As for WSD-IA, additional infrared touch frames were put in place, increasing the number of touch enabled screens to eleven. 
As text input with the virtual keyboard on WSD-VW was quite cumbersome, and to ensure a fair comparison between the conditions between both WSDs, participants were also provided with a physical keyboard for the touch condition, but its use was limited to text input. We have also provided a height-adjustable movable table with WSD-VW, to allow users to put down the interaction device anywhere in the room (see Figure~\ref{fig:system2}).

\begin{table}[htbp]
\caption{Order of test conditions for the second user study using Miro.}
\label{tab:testConditions2}%
\begin{tabular}{lll}
\textbf{Condition ID}     & \textbf{WSD}   & \textbf{Interaction method}  \\
\hline
\multicolumn{3}{l}{\textbf{Participant 1 (p1)}} \\
\hline
IA-Keyboard-p1   & WSD-IA    & Keyboard             \\
IA-Tablet-p1     & WSD-IA    & Tablet               \\
IA-Touch-p1      & WSD-IA    & Touch                \\
VW-Touch-p1      & WSD-VW    & Touch                \\
VW-Keyboard-p1   & WSD-VW    & Keyboard             \\
VW-Tablet-p1     & WSD-VW    & Tablet               \\
\hline
\multicolumn{3}{l}{\textbf{Participant 2 (p2)}} \\
\hline
IA-Tablet-p2     & WSD-IA    & Tablet               \\
IA-Touch-p2      & WSD-IA    & Touch                \\
IA-Keyboard-p2   & WSD-IA    & Keyboard             \\
VW-Keyboard-p2   & WSD-VW    & Keyboard             \\
VW-Touch-p2      & WSD-VW    & Touch                \\
VW-Tablet-p2     & WSD-VW    & Tablet              \\
\hline
\end{tabular}
\end{table}

The \textbf{task} consisted in using Miro to replicate the same UI as in the first study (see Figure~\ref{fig:task}). 
The participants first discovered Miro on a desktop computer for two hours, with the goal of reproducing another small UI similar to the main one in terms of components to be drawn. Detailed task, protocol, and discussion guide are available in supplementary material~\cite{annex2}.

We used similar interaction methods as before, i.e., a wireless keyboard with a touchpad, touch on the WSD with a keyboard, and a tablet synchronized with the WSD (see Figure~\ref{fig:system2}). We changed the order in which participants used the interaction methods, as shown in Table~\ref{tab:testConditions2}. 
For logistic reasons, the tests were run on WSD-IA first, and then on WSD-VW.

%%TODO explain the difference between AI and VW mockup is inspired of a true interface developped on both WSDs. The resolution was keeping the same, but the size of the pixels is not the same. As it was noted more convenient to not adapt the size of the interface on VW when the interface was developped, it was keep like that.

%\subsubsection{Observation methods}
%We relied on the same recording setup as for the first study for the WSD-IA, using the same video cameras and microphones to record the sessions: top, front and back cameras for WSD-IA. %, and back, left, and right cameras for WSD-VW. % For the WSD-IA, three video cameras were used, at the top, front (middle of the WSD) and back (at top of the opening). 
%For the WSD-VW, three cameras were used at the back, and positioned at each ends of the WSD.

%We tracked also the participants movements in the WSDs space using a tracking pipe based on Kinect and \(/\)psi, see~\cite{cdve-tech}.
%Comments and actions were analyzed thematically to identify encountered issues. 

At the end of each session, the facilitator debriefed with the participants to discuss their feedback and suggestions for improvement. The debriefing collected spontaneous open-ended feedback first, and then leveraged sentence completion~\cite{lallemand2015methodes} regarding first impression, the utility, usability, and user experience of the test condition, as well as its capacity to meet the needs and desired improvements. 
See the conversion guide and user study guide provided in supplementary material~\cite{annex2}.% a debriefing was done to collect feedback of participants and improvement suggestions (see the conversation guide available in Annexe).
% The user study guide is available in annex~\cite{annex2}.

\subsection{Results}
%designing 1:1 scale
Designing in real size was appreciated by participants, it enhances the understanding of object dimensions, placement, and the use of space (``\emph{the model on a large scale gives more precise impression of the objects' size, of their position and the use of space.}'' p1, ``{It is great to see the result directly in real size on the targeted display.}'' p2).

\subsubsection{General observations}
\begin{figure}
    \centering
    \includegraphics[width=0.99\linewidth]{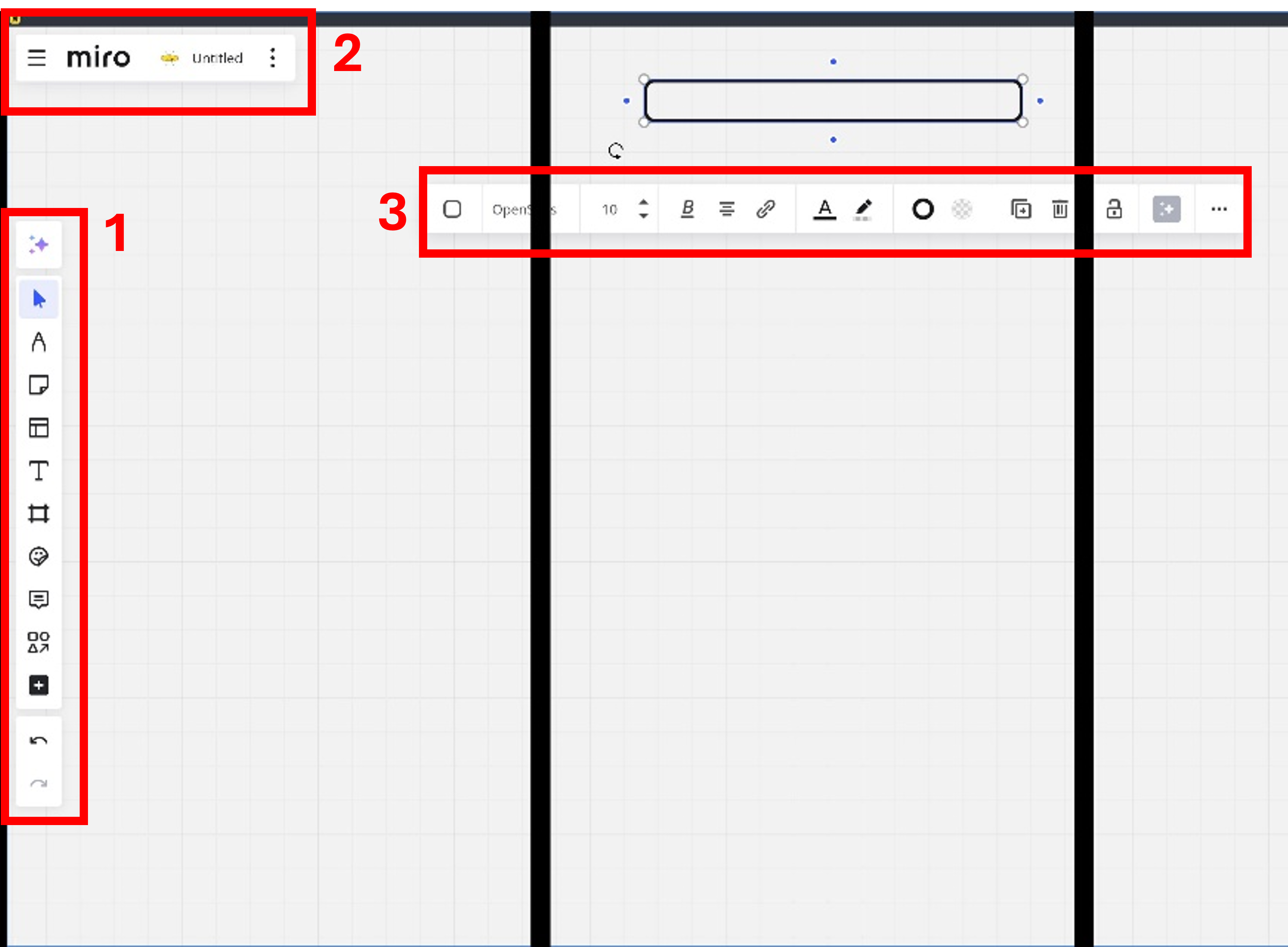}
    \caption{1)~shows the left menu used to create objects, 2)~is the menu used to set some parameters for the workspace or go back to the list of projects in Miro, and 3)~is the context menu of the selected object.}
    \label{fig:miro}
\end{figure}

\paragraph{General observation about Miro}
%%menus position
In Miro, the main menu that allows to create objects has a fixed position on the left side of the workspace and is centered vertically. The context menu for a selected object is displayed in the vicinity of the object, as shown in Figure~\ref{fig:miro}.
Nobody complained about the position of the two menus. The context menu allowed to modify properties and perform actions on the selected object, such as deleting or duplicating it, while the main menu on the middle-left side allowed the creation of objects (``\emph{It is convenient that a part of the functionalities [object properties] are placed in the working zone.}'' p2). The only exception to the above is for p2 in condition touch on WSD-IA (``\emph{It's not working well.}'' p2, ``\emph{It's not comfortable.}'' p2). We however note that the mock-up to reproduce mainly used the leftmost parts of the WSD. Had UI elements from the mock-up been placed in a more central position (or even towards the right side of the WSD), it is likely that the position of the main menu would have been more of an issue. The lack of complaints on the menu placement could also be due to the fact that the mock-up only contained a moderate number of objects that the participants must create. Another explanation could come from a strategy used by both participants to mainly rely on duplicating an existing object then completely modifying its properties, rather than creating a new object using the menu. This behavior could also be a reaction to the problematic placement of the menu.
The context menu was used a lot, even though it sometimes stopped working for no apparent reason. %(in that case, the web page was refreshed to circumvent the issue).% It was sometimes broken for no reason.% 
The top menu, which allows users to set some parameters of the workspace, was only used to change the background color of the workspace. 

%undo redo: not necessary ? Except one time when objects were deleted accidentaly

%Miro issues
The most critical aspect of Miro noted by both participants is that it does not provide a rich objects library. There are for example no chart elements (``\emph{The available objects are very basic, so everything must be laboriously drawn by oneself.}'' p1). %It seemed not possible to chain two actions in a row on the same object (such as changing the text size and then its color) with the touch and tablet modalities. It is then necessary to deselect the object by clicking elsewhere, then reselect the object ("When an object was already selected and a first action had been performed, the contextual menu was hidden and therefore, I could no longer perform another action. I then had to click next to the object and then reselect it." VW-Tab-p1).
Both participants complained that the minimum font size is blocked at 10. 
It was observed that, when an object was created (or duplicated) partially outside the current viewport (the portion of the workspace that is visible on the WSD), then Miro adapted the view to include that and all the objects (``\emph{The zoom level changes when I copy/paste something out of the view, to put everything in view.}'' p2). This disturbed the participants.
During the session on WSD-VW with p1 and the touch condition, a modal popup window opened in the middle of the WSD, blocking all interactions. It took p1 a whole minute to notice it, which was frustrating. 
% He was meanwhile frustrated for not being able to interact anymore (the modal popup window was blocking all interactions).
Both users liked the ease of changing the background color of the workspace.

\paragraph{Participants' strategy}
The participants often used a strategy consisting in subdividing the mock-up in panels or zones, using a lot of copy and paste actions, and working on main structural elements before tackling the details, for all conditions. With both the touch and the keyboard conditions, p1 started by creating objects at the center level, to interact at comfortable height. He then panned the mock-up to move the objects up. P2 also used this strategy but only when interacting with touch (``\emph{It is practical to be able to pan the mock-up, because I was able to work in the middle of the screen at the beginning, and then move everything up.}'' p2).

\paragraph{Task duration}
One-hour sessions were judged acceptable, but chairs were deemed mandatory for longer sessions.

%moves [Adrien]
%WSD-IA asked less moves.
%should be: a lot with touch, a little with the keyboard, almost nothing with the tablet.
%should be more in the VW than in the IA
%"My calves hurt less than with the keyboard, maybe it's because I didn't stamp as much." p1

\subsubsection{Interacting with a wireless keyboard equipped with a touchpad}
\begin{figure}[t]
    \centering
    \includegraphics[width=0.99\linewidth]{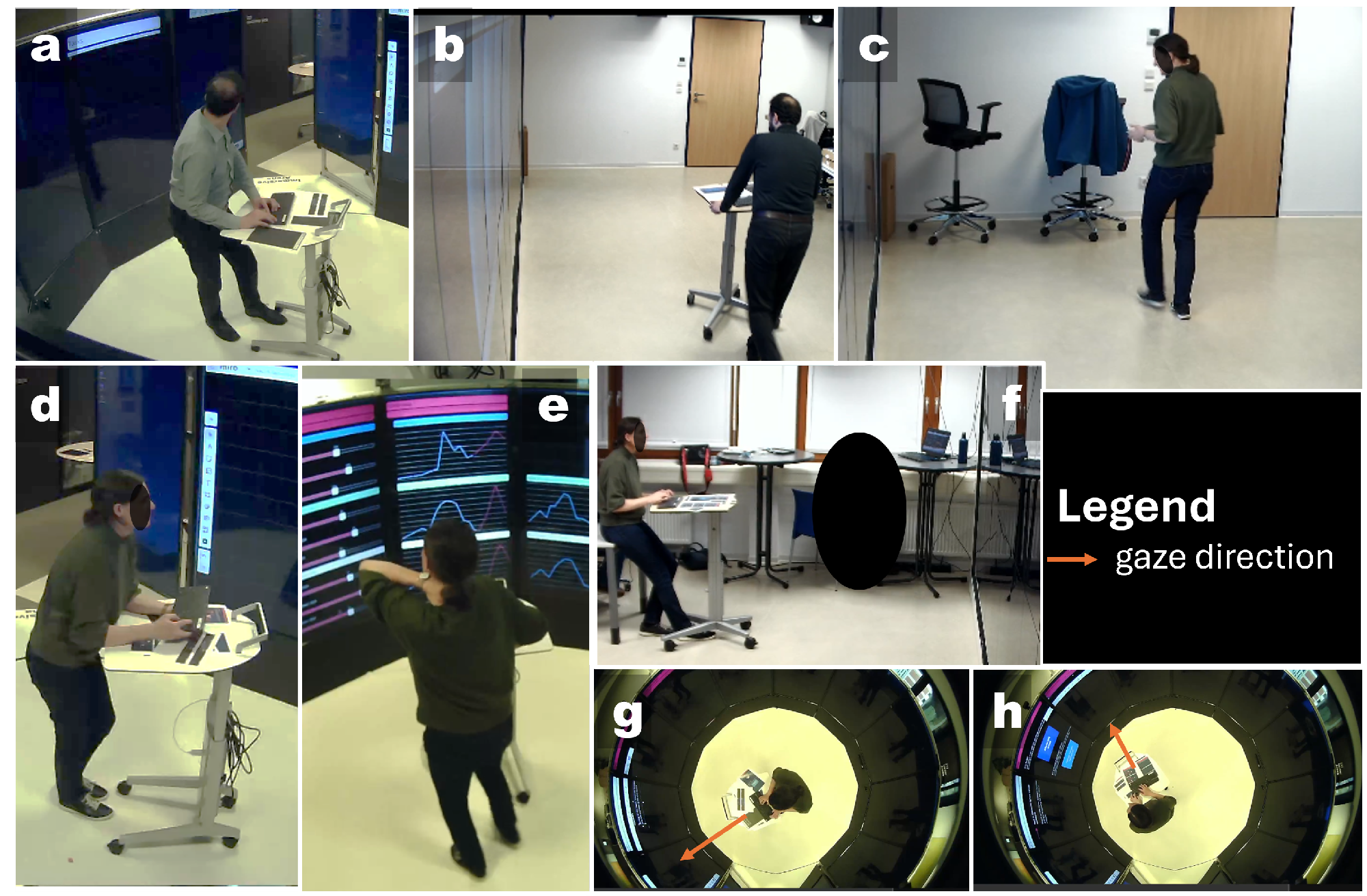}
    \caption{Using Miro with keyboard+touchpad. a) P1 turns his torso to look behind in WSD-IA. b) P1 moves the table with the keyboard on in WSD-VW. c) P2 carries the keyboard in her hands in WSD-VW. d) P2 leans on the table and tilts the keyboard with her hands in WSD-IA. e) P2 massages her neck in WSD-IA. f) P2 sits down in WSD-VW. g) and h) p2 turns the table to face the work area.}
    \label{fig:keyboardXP2}
\end{figure}

%global strategies
    %moves
%AI p1
In WSD-IA, p1 left the keyboard on the table and turned his head and torso to see what he was doing, particularly when he created the last panel of the mock-up, which was behind his back, see Figure~\ref{fig:keyboardXP2}.a. Once, he turned around the table without moving it.
%VW p1 
In WSD-VW, p1 always left the keyboard on the table including when he moved (five times) to the extreme right to change the zoom level from the zoom menu (at the opposite end of the WSD, in the bottom right corner) or to create the last panel of the mock-up. In this case, p1 rolled the table until he was facing the zoom menu or the new work area (see Figure~\ref{fig:keyboardXP2}.b).
%WSD-IA p2
P2 noted that she moved more using the keyboard with WSD-VW than with WSD-IA. %oui mais pas tellement plus puisque qu'elle s'est assise
In WSD-IA, p2 remained at the table and turned it to face the work area (see Figure~\ref{fig:keyboardXP2}.g and h). P2 mainly leaned over the table (see Figure~\ref{fig:keyboardXP2}.d), but she often held the keyboard on her arm during the last half of the session. P2 kept the keyboard in her hands most of the time when doing modifications like creating an object, moving or resizing it, changing its properties or zooming and panning. To enter text, p2 mostly put the keyboard on the table. She showed some signs of physical fatigue halfway through the session (movements with the neck, massages of the neck, see Figure~\ref{fig:keyboardXP2}.e.
%VW p2
In WSD-VW, p2 moved the table with the keyboard on it to face the panel she was working on, and halfway through the session, sat down for 15 minutes (see Figure~\ref{fig:keyboardXP2}.f). But p2 also often stood up while holding the keyboard in her hands. Occasionally, p2 stepped back to get an overview. 

    %strategy
%p1 WSD-IA
In WSD-IA, p1 started by creating the left panel (Tasks), then he created the title and subtitle of each panel. He looked for some chart objects or chart icons. As he did not find any, he instead went on to focus on creating the sliders. 
%p1 VW
In WSD-VW, p1 adopted another strategy. He first created one big blue rectangle to serve as background for all panels. Then, he created the other objects on top of it. This choice generated difficulties later on, when trying to select objects without selecting the background shape. He created the titles in the middle of the WSD, then panned the workspace to place them at the correct height. He built the interface panel by panel from the left to the right. He sometimes zoomed in to work on details (e.g., icons). He had some trouble grouping and ungrouping objects. 
%p2 WSD-IA
In WSD-IA, p2 started to design objects directly where they were supposed to be placed. After creating the titles, p2 designed panel by panel from left to right, before coming back to the charts and sliders to add some details. At the end, she added blue boxes to create the background of each panel. 
%p2 WV
In WSD-VW, p2 created the objects at a lower position than their final placement, then panned the workspace to adjust the height. After first creating the titles, she then designed the remaining objects panel by panel from left to right. 

%generalities
P2 used keyboard shortcuts a lot, to delete, copy, paste, but also to create new shapes, text boxes, or zoom in and out. P2 and p1 used keyboard arrow keys to place objects.
%Only p2 used one time the zoom menu at the bottom right (in IA).
%%10:41 La taille de l'objet créé semble relatif à la taille de l'objet créé précédemment VW-p1
%P1 and p2 created the objects at the center (vertically) of the WSD and then panned the mockup to place it correctly at the top of WSDs. 
%positive
This facilitated a precise placement of objects (``\emph{I tried to be accurate with the position of the objects from the paper mock-up, taking into account the position relative to the various screens of the wall.}'' p1). It also allowed participants to be more precise in the layout of the mock-up itself relative to the workspace. (``\emph{With the keyboard, positioning the scene correctly is easier.}'' p1). 
The keyboard was also appreciated for entering text. P2 also found it easier to draw graphs curves using the keyboard touchpad than with the tablet or in the touch condition.

%negative
The participants disliked the touchpad because it lacked precision (``\emph{It was OK with the touchpad, but for smaller things it would be better with a mouse.}'' p2). P1 complained about triggering actions inadvertently, such as an unintended pinch leading to a change of zoom, but unwanted panning, and selecting/deselecting actions also occurred (``\emph{Of all the modalities tested, the keyboard was the most frustrating one, due to the touchpad's extreme sensitivity. As a result, this modality is difficult to use alone, it would be better to complement it with other modalities.}'' p1). That's why participants asked for an option to change the zoom level independently of the panning. P1 and p2 in general prefer to use a mouse instead of the touchpad. It however remains unclear how a mouse would be dealt with when a participant walks around with the interactive device in hand, since a mouse might be difficult to use without being placed on the table. P2 also used some shortcuts to avoid using the touchpad, such as those enabling the creation of a new shape or a new text box at the cursor's position (``\emph{But it's not nice with the touchpad, so I tried to avoid using it, and used shortcuts instead.}'' p2).
The participants were uncomfortable when inputting text (``\emph{The table height was too low, which hurt my wrist. I tried to raise it, but it was already at the maximum height. In the end, I carried the keyboard to be more mobile, more flexible and to avoid wrist pain.}'' p2)
The workspace moved a lot inadvertently for both participants, who wished they could lock the zooming and panning.

\subsubsection{Interacting using direct touch on the WSD}
\begin{figure}
    \centering
    \includegraphics[width=0.95\linewidth]{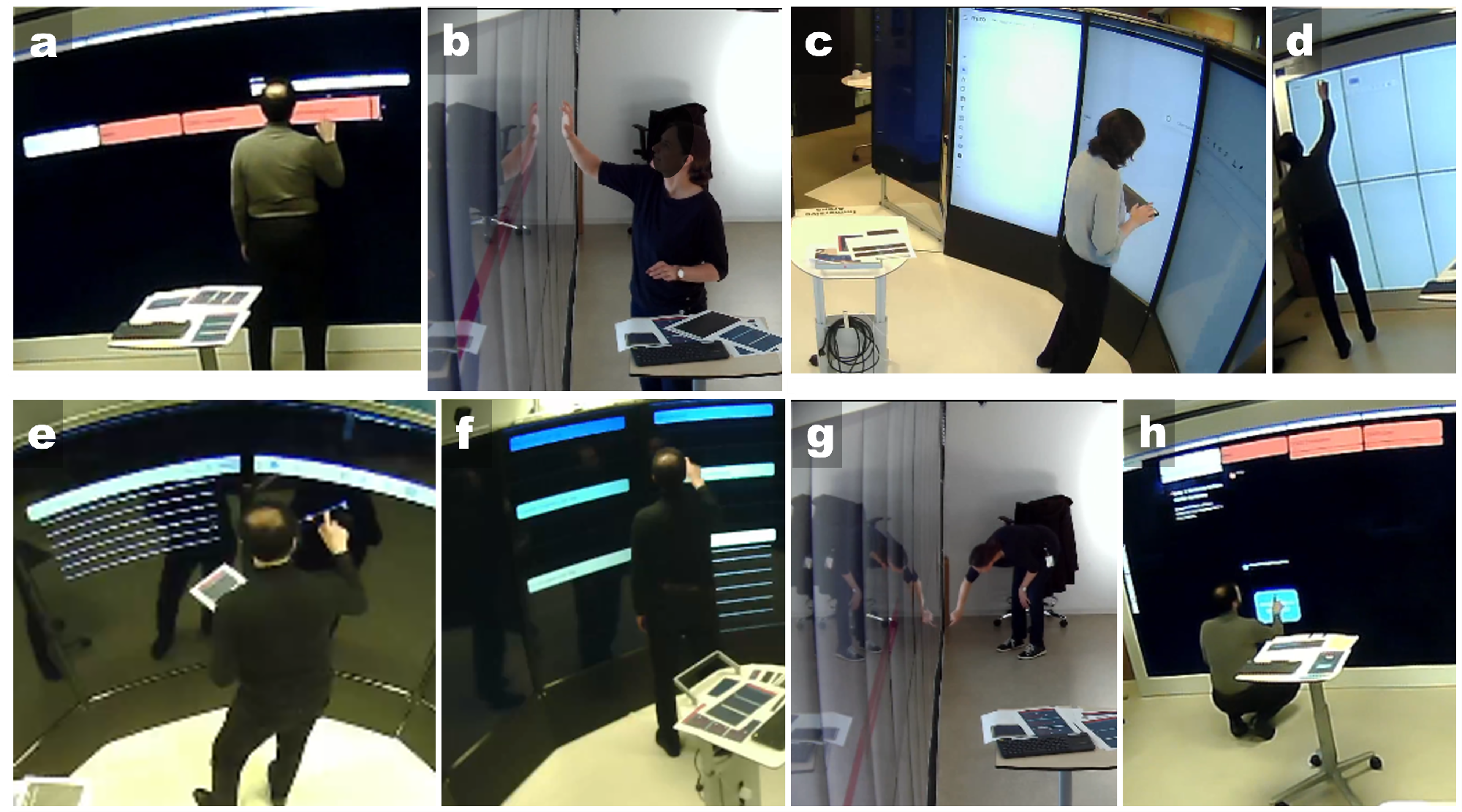}
    \caption{Using Miro with touch. In a) b) and c) participants create the objects at the WSD center. e) shows p1 zooming in and panning to design details, and then in f) repositioning the workspace correctly. In a) and f) p1 leaves the table in the middle. In b) p2 puts the table closer. d), g) and h) show moments when the participants interact too low or too high.}
    \label{fig:touchXP2}
\end{figure}

%global strategies
%moves
P1 preferred not to move the table. He mainly left the paper mock-up and the keyboard on the table (``\emph{I didn't move the table, I preferred to have the papers in my hand in front of the display when I really need them. I didn't want to be cluttered with the table.}'' p1). P2 preferred to move the table closer to the WSD to let the paper mock-up on it and to have quick access to the keyboard when needed, while maintaining access to the left menu on the WSD. She often moved the table to follow her process of working from left to right, but also sometimes pushed the table out of her way (see Figure~\ref{fig:touchXP2}.b).
We also observed that both participants stepped back from time to time to get an overview (both participants did so for WSD-VW and only p2 for WSD-IA; ``\emph{We can't check the exact position of the created objects; to do this we need to move back.}'' p2).
They both knelt down a few times in WSD-VW to interact with the zoom menu, and p1 also did so to quickly modify some objects (``\emph{I had to kneel down, that was more physically demanding than with other modalities.}'' p1, see Figure~\ref{fig:touchXP2} g and h).
Touch was the modality requiring the most movement. Some physical fatigue arose, particularly in WSD-VW for p2 where she worked closer to the WSD (``\emph{It hurts my neck when I am so close to the display and I look up. And it is warmer when we are close to the screens.}'' p2).
%P2 mainly placed the table to her right, close to the WSD, to have easy and quick access to the paper model and the keyboard while leaving free access to the left menu on the WSD.

%strategy
We also observed that p1 and p2 first created the top objects of the mock-up at arm's height (vertically centered on the WSD), then moved the workspace upwards to correctly place the different elements ``\emph{I moved [panned] the designed interface so that it was at my height.}'' p1, see Figure~\ref{fig:touchXP2} a, b and c). %("When I worked higher it is tiring, like for positioning the sliders"
On WSD-VW, both participants made use of the zoom in and out capabilities several times to design the smallest objects and then return to the initial view (see Figure~\ref{fig:touchXP2} e then f). P2 noted that it was tiring to work at the top or the bottom of the WSD, especially when using the zoom widget at the bottom right (see Figure~\ref{fig:touchXP2}.g).
They both created the titles first and then worked panel by panel. This eased the interactions and decreased the difficulty of working on such a huge surface by staying in a limited area and decreasing the amount of physical movements (``\emph{I designed panel by panel, so I had the feeling of working only locally.}'' p1).%Then, I didn't need to have a global view

%VW vs IA
%reculer pour mieux voir
A notable difference between WSD-IA and WSD-VW is that the participants moved back more often and further back, to get a bigger picture with WSD-VW than with WSD-IA (``\emph{I needed to move back to have a global view.}'' p1, when interacting on WSD-VW). 
%better touch on VW
It was also noted by p2 that touch on WSD-VW was more tiring than on WSD-IA. This could be explained by the fact that in WSD-VW, the zoom widget was far from the users when they stood in the main working area. To use it, they were forced to kneel down and press buttons, which is uncomfortable. However, they used the zoom widget more often on WSD-VW than on WSD-IA (only one time on WSD-IA, by p2), which might be due to the issues encountered with touch-based zooming and panning actions, which were sometimes too fast and sometimes laggy (``\emph{It is more tiring than with WSD-IA because of the zoom button at the bottom right.}'' p2). 
Overall, both participants had the feeling that touch worked better on WSD-VW than on WSD-IA.
%positions bottoms and top not comfortable to interfact
%Furthermore,  
%P2 noted that it is tiring to work at the top and bottom of the WSD using touch, although this was less of an issue in Miro where the menu used to create objects is placed on the left, rather than the top of the window in Figma. % ("It is tiring to work at the top or the bottom of the the WSD." p2).

%positive
The touch modality helps the most with the sense of space. It is also more instinctive and rapidly allows to place the created objects near their target location, even if the positioning is rarely immediately perfect (``\emph{For some actions it is nice to do them directly on the display, e.g., when I add a text box, I can place it directly where I want.}'' p2). 
The feeling of working directly on the interface was liked by participants (``\emph{We can change directly on the display.}'' p2).

%negative
%not reliable technology
Many issues appeared with touch on both WSDs. The sensitivity of the touch-based gestures was not adapted; sometimes it was too sensitive, sometimes not responsive enough. 
%p1: "The attention needs to be shared between the paper mockup and the screen when text is entered. So, I was not able to see directly if a typo was done."
The technology used to recognize touch events was questioned (``\emph{The touch needs to be improved.}'' p1). Indeed, participants wondered whether the unexpected zooms, pans, and deselections/selections were triggered by pre-touch when a finger, sheet of paper or sleeve inadvertently entered the infrared detection zone of the touch frames (e.g., pointing a finger while thinking ``\emph{I wonder if it is due to a finger entering the detection zone of the infrared frame.}'' p1, ``\emph{being very close to the display is sometimes enough to be detected as a touch.}'' p1, ``\emph{It was disturbing when my pointing finger was close to the surface, because I was thinking before acting and then a touch was detected.}'' p1). %Maybe it is due to the way Miro manages touch events, which was designed for desktop- and tablet-based interactions. This might also be due to the infrared frames not being precise enough, or being too sensitive.
%manipulation
Both participants had trouble placing and resizing objects, particularly for the first object with this modality %. 
%The participants noted that it was not enough easy to resize and position exactly the objects using touch 
(``\emph{I still have some issues with the touch behavior, notably the resizing of the objects and text fields.}'' p1). This was particularly the case when objects were tiny (``\emph{When the manipulated objects are too small, it is impossible or very hard to modify them, like with the sliders or the curves.}'' p2). Furthermore, grasping tiny objects such as the handles used for resizing objects is complex and sometimes even impossible (p1 \& p2). Modifying tiny objects therefore requires zooming in, and it then becomes difficult to readjust to the right zooming level and to position the mock-up in the workspace afterwards (``\emph{When objects are very small, like the one composing the video camera icon, it was complicated to position them. So, I zoomed in. Then, I used the zoom menu to go back to a 100\% zoom level.}'' p2). Hence, the participants asked for an option to restore to the initial view, or at least to be able to go back to a given fixed view of the workspace.

%layering
Positioning the objects on different layers (z-index) was also problematic. Ensuring that each object was in the right layer was sometimes tricky. But, the main issue was that, when participants wanted to select an object in the foreground, a background object was often selected instead (``\emph{When we are drawing objects composed of simple forms, it is complicated to specify what should be in the front/background.}'' p1). 

%unexpected zoom & pans
The amount of unexpected zooming, panning, and deselecting/selecting actions hampered work progress. In reaction, both participants had to frequently zoom and pan. They asked for the possibility to perfectly re-center the mock-up and resize the workspace based on the WSD dimensions (``\emph{I need a button to rescale the scene.}'' p1).

\begin{figure}
    \centering\includegraphics[width=0.99\linewidth]{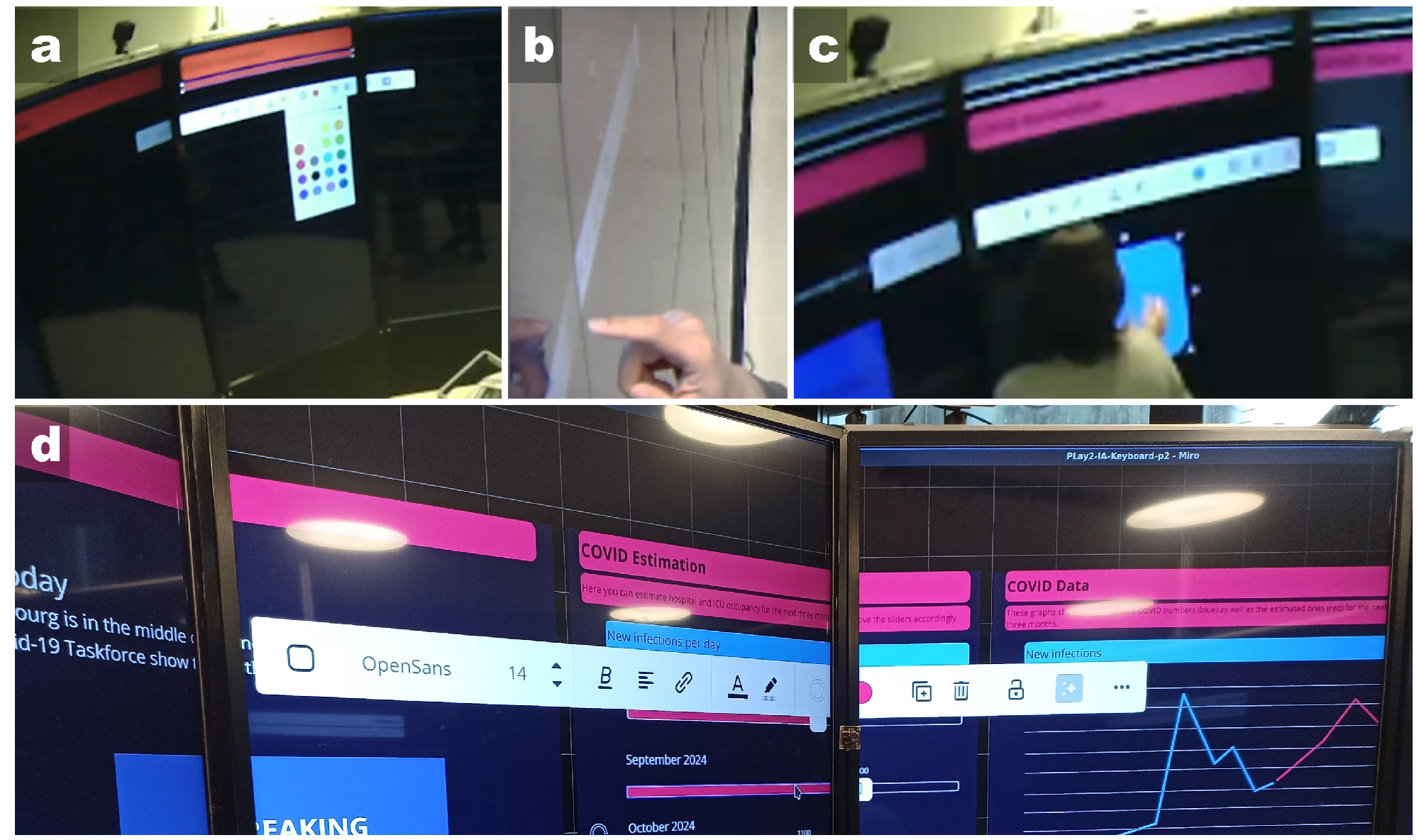}
    \caption{Bezel issue on WSD-IA (a, c and d), and on WSD WSD-VW (b).}
    \label{fig:bezel}
\end{figure}

%bezels
Screen bezels between the tiles composing the WSDs made the touch interaction harder. It was difficult to drag an element across screens because the bezel in between would interrupt the gesture (``\emph{Objects need to be large enough to be moved across screens. Smaller bezels between two screens would improve the touch reliability.}'' p1). %Drag and drop were sometimes perturbed by the bezels  %And objects need to be enough large to be draggable from one screen to another, due of the bezels. 
Sometimes, the bezel occluded partially the options in the context menu (see Figure~\ref{fig:bezel}). Participants worked around it by first moving the selected object to ensure that the whole context menu was visible, then repositioned the object in the right position. 
Bezel-related issues were more raised by the participants with WSD-IA than with WSD-VW, likely because they are wider for that WSD.

%straigh lines
Drawing straight lines was not possible using touch (p1 \& p2 ``\emph{Drawing a straight line is complicated with touch because of finger tremor.}'' p1).

%keyboard
While typing with the keyboard was appreciated, using an additional device poses a classic parking problem or requires the user to carry it around (``\emph{The physical keyboard is better than the virtual keyboard on the tablet, and I was able to input more text. But the fact that the keyboard is placed on the table is not comfortable. Why not searching for a solution to have the keyboard closer without having to hold it in the hands, like attaching it to the left arm?}'' p2).

%At the end, the positioning (on the surface and on layers) and resizing of objects was complex with the touch modality.

\subsubsection{Interacting on a synchronized tablet}
\begin{figure}
    \centering    \includegraphics[width=0.95\linewidth]{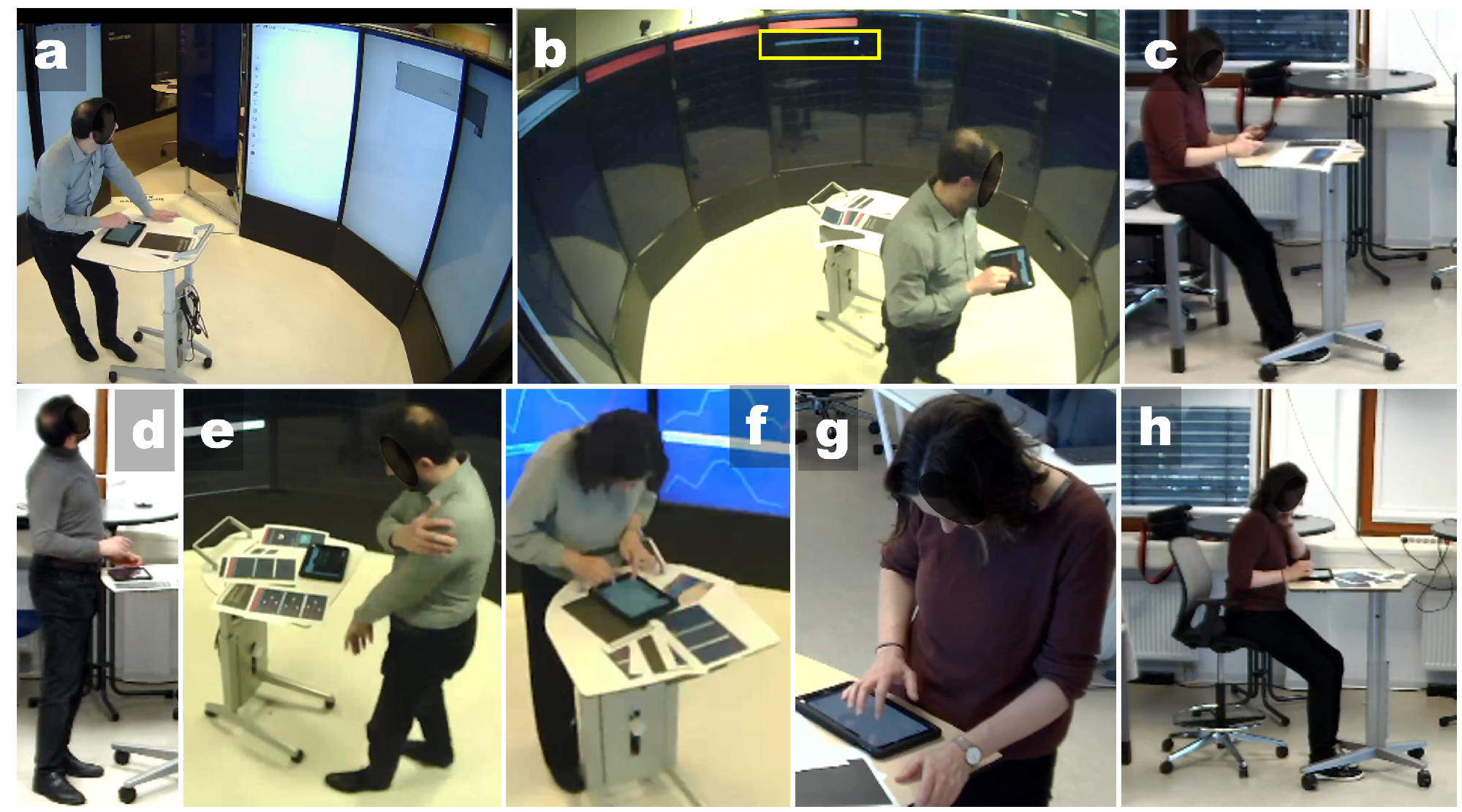}
    \caption{Using Miro with a tablet. a) and g) participants placed the tablet on the table and their non-interactive hand, b) p1 holds the tablet in hands, c) and h) shows p2 sitting down on a table and on a chair, d) and e) shows that p1 is uncomfortable and stretches f) shows p2 entering some text with two hands with the tablet placed on the table, in b) the yellow rectangle shows the object manipulated by p1 that is in his back.}
    \label{fig:tabletXP2}
\end{figure}

%I wanted to move the guidelines on the wall to align them of the structure of the screen composing it. It feasible in Miro , but was not know by the user (solution --> changing the arrow to the hand cursor, not a Miro issue).
% 10:03 Not same icons on tablet and on the wall? VW-p1
%global strategies
%to check if they interact on the tablet then check on the wall
%check if colors are really different
%moves
Both participants moved very little during their sessions with the tablet.
In WSD-IA, p1 and p2 stayed at the table and mostly did not move it. They both held the tablet during about half of the session, but p1 did so in the first part (see Figure~\ref{fig:tabletXP2}.b), while p2 rather did that in the second half. They both changed their holding strategy because of the pain they felt in their arm (p1, see Figure~\ref{fig:tabletXP2}.e) or neck (p2). They both showed signs of physical fatigue in the second half of the session (see Figure~\ref{fig:tabletXP2}.d). 
P1 started the WSD-IA session with the tablet in his hand, then put the tablet on the table, complaining about the weight of the tablet. After a while, since the position at the table was not comfortable either, p1 took the tablet back in his hand.
For WSD-VW, p1 left the tablet on the table throughout the session.
None of the participants moved the table extensively, and they did always face the area they were working on. For instance, p1 created the rightmost panel on WSD-VW while staying at the table that was placed at the left of the WSD, as shown on Figure~\ref{fig:tabletXP2}.b.
P1 also sometimes held the tablet to move objects (e.g., title and subtitle of the rightmost panel) from a screen in his back to a target position he was looking at, without gazing at the objects during their movement. When the tablet was placed on the table, they both put their other hand (the one they were not using for interaction) on the table (as shown in Figure~\ref{fig:tabletXP2}.a and g), except for p2 when she was entering text and therefore needed both hands (see Figure~\ref{fig:tabletXP2}.f). 
P2 changed the tablet orientation from landscape to portrait in both WSD environments, for the whole second half of the session for WSD-IA and during 15 minutes in the beginning of the second half of the session for WSD-VW, before switching back to landscape in that case. 
P2 often tilted the tablet.%, with one or two hands.
%p2 started the session (IA) with the tablet on the table, then p2 chose to hold the tablet because of neck pain, even though p2 remained standing at the table. 
During the WSD-VW session, p2 sat down on a table for the first fifteen minutes of the session (see Figure~\ref{fig:tabletXP2}.c), then on a chair until the end of the session (see Figure~\ref{fig:tabletXP2}.h). 
With WSD-VW, they both moved a little to check results on the WSD, leaving the tablet on the table.
They both looked at the tablet for the creation and modification of objects, and then regularly checked the results on the WSD. As for the placement of objects, they partially looked at the tablet, mainly when the titles had already been created (as they then had a reference), but also looked at the WSD, although the lag between updates on the tablet and their effect on the WSD impeded these checks. They both opted to directly position the objects precisely on the WSD. 
They both also worked with the tablet by zooming in on part of the working area.

%On the VW, p2 created the objects on the tablet, then checked on the VW the result. But for the movement of the created object, p2 was looking directly at the WSD, although the lag made this verification difficult. P1, when interacting on the VW was looking at the VW, when placing the objects.

%%Positive
The most interesting aspect of this modality is that the viewports of the tablet and that of the WSD are dissociated. This allows the user to zoom in considerably on the tablet while keeping the view on the WSD unchanged. This was noted as the main advantage of the tablet modality by both participants (``\emph{Dissociating the viewport of the tablet and that of the wall, especially for the zoom and pan}'' was pointed out as the reason that makes the system easy to use by p1, ``\emph{The decoupling between zoom and pan on the tablet and on the wall is interesting. As the tablet is tiny, I needed to zoom a lot.}'' p1 ``\emph{The workspace on WSD-VW never moved, this is really comfortable.}'' p2). 

This zoom in also helped to place elements more precisely (as touch is more reliable on the tablet than on the WSD) (%p1,
``\emph{With zoom [on the tablet], it becomes possible to position smaller elements.}'' p2). The ability to zoom in considerably cancels out the fat finger effect observed with the touch modality (p1)~\cite{voida2009getting}.
Creating small objects was experienced differently by each participant. For p1, the ability to zoom on the tablet without affecting the view on the WSD enabled him to work on the smaller objects and their details. However, p2 still found the tablet inadequate to design tiny objects (``\emph{I avoided working on the very small details with the tablet, it was not adapted.}'' p2)

P2 appreciated a lot the tablet condition, particularly on WSD-VW (``\emph{It was with the tablet that I was able to do the most. It works best.}'' p2).

%Negative
The most negative aspect regarding the use of the tablet was the delay between actions performed on the tablet and their effect becoming visible on the WSD (p1 \& p2, ``\emph{The tablet seemed slow, it didn't respond right away.'' p2, ``There is a delay between moving something on the tablet and seeing it move on the WSD.}'' p2). P1 additionally observed disconnections on the tablet at times.

Both participants found the tablet rather heavy, which made it difficult to carry it around. %%Already noted in the first XP.
They therefore put it on the table, but the ergonomics of the table were not good enough, which led to pain in their wrist, neck, back and shoulders. The pain levels reported by the participants were more intense with WSD-VW. We note that the table was not exactly the same as that of WSD-IA, which could explain the increase in complaints (``\emph{The tablet is heavy to carry. So, carrying it in my hands didn't suit me. So, I put it on the table. But in this position, my shoulder and neck muscles were hurting. I should have raised the table up, but I didn't think about it.}'' p1, ``\emph{I had a bit of pain in my wrist due to the position while interacting, especially while entering text.}'' p2).

Positioning the objects precisely with the tablet was complicated (``\emph{You can't place [the objects] as precisely as you would like.}'' p1). 
Grouping objects was not easy either, using touch input on the small surface of the tablet.
The guiding and snapping lines to help with aligning objects were visible only when interacting on the tablet. P1 would have preferred to see them on the WSD as well.

The working surface on the tablet is small, especially when the virtual keyboard is opened (``\emph{But it is very small: the visible surface is very small, and the virtual keyboard takes a lot of the available space.}'' p2). For this reason, p2 changed the tablet's orientation from landscape to portrait, to see the workspace better when the virtual keyboard was opened.

Participants deemed drag and drop difficult on the tablet, so was multiple selection. The drag and drop of an object was in fact accelerated depending on the distance on the tablet. However, as the surface is small, the acceleration quickly becomes too pronounced, which disturbed p1 (``\emph{When I moved an object, it seemed like the speed of moving was changing depending on the distance.}'' p1). 

It was also noted by p2 that the available colors on the WSD and the tablet seemed slightly different, making the selection of the right color harder (``\emph{It was harder for me to find the right colors. Maybe it's because they are not exactly the same on the tablet and on WSD-VW, or perhaps they are not perceived the same way.}'' p2).

Last but not least, both participants did not like to type text with the tablet; they preferred the physical keyboard.
The predictive text input of the tablet's virtual keyboard was not working well. Indeed, 
after p1 copied a text box, deleted the text in the new text box, and started to input new text, the predictive text input proposed a word starting with the first letter of the previous text %, that is well no more visible in the textbox. 
(``\emph{It's a weird behavior, and it poses a problem with the strategy I used for copying/pasting and then modifying.}'' p1, ``\emph{Typing is more painful than with a keyboard. But with a more efficient automatic text completion system, it would be faster. But, here, it is not efficient enough. When I duplicated a text box and then deleted its text to enter new text, it seemed that the system was keeping the first letter in mind, although it was deleted, as it was still considered by the text completion system.}'' p1).

%But when I released the selected object, it happened that the object was moved a little bit (for instance when I placed the subtitle texts under the title text. VW-tabl-p1

To improve the use of the tablet, both participants suggested using a stylus for more precision, and a keyboard, mainly to facilitate text entry (``\emph{A stylus to better move the small elements and a physical keyboard.}'' p2). They also suggested using a wider tablet and would have liked a more ergonomic table.

\subsubsection{Suggested improvements}
%improvement propositions
Participants proposed different features and options to improve the 1:1 scale design on a WSD.
\begin{itemize}
    \item an option to lock and unlock zoom on WSD (p1 \& p2);
    \item an option to lock and unlock panning on WSD (p1 \& p2);
    \item a button to center \& zoom the design within the WSD viewport (p1 \& p2);
    \item an enriched objects' library (p1 \& p2);
    \item an improved touch detection system (p1 \& p2);
    \item more flexibility regarding the font size (p1 \& p2);
    \item the ability to move the context menu to work around occlusions by the bezel (p1 \& p2);
    \item freehand drawing with an AI-based conversion of hand drawings into standard UI elements (``\emph{It would be interesting to be able to draw by hand, to make sketches.}'' p1, ``\emph{Hand drawing with an AI that recognizes drawings and normalizes drawn shapes.}'' p1, ``\emph{Some AI assistance, which gets what I'm doing and enables me to go faster.}'' p1);
    \item speech recognition to dictate text and for simple commands (``\emph{As well as using some vocal commands, for instance to duplicate an object or change the color}'' p1);
    \item limiting the number of sub-menus, particularly in the context menu, to 2 or 3 at most (p1);
    \item limiting pop-ups as much as possible; those that cannot be avoided should not appear in the middle-center part of the screen, but instead appear in the work area. (p1);
    \item a smart feature that recognizes series of actions that are repeated, to ask the user whether and how many times to repeat them (``\emph{a smart function that, for example, if we repeat the same gesture [action] several times, then the system asks if we should repeat it again and how many times, e.g., for the labeled ticks along sliders.}'' p1);
    \item a wearable keyboard (p2);
    \item supporting mid-air gestures (``\emph{And maybe find a solution to move less [than with touch], like using mid-air pointing instead of touch, to support working from a distance.}'' p2)
    \item adding specific actions for each tile composing the WSD, such as changing that screen's background (``\emph{It would be interesting to be able to change the background for each screen}'' p2), that would require to declare the tiling configuration of the WSD.
\end{itemize}

%\subsection{Discussion}
\section{Discussion}\label{sec:discussion}
%\subsection{Main findings}
In comparison to the first user study, the sessions of the second one lasted longer (the full hour in all cases) and participants managed to advance much further in the creation of the mock-up. Hence, we can say that with Miro, participants could work more efficiently, and this tool can be considered as better adapted for the design in 1:1 scale on WSDs than Figma. Our observations have shown that this improvement was mainly due to the main menu being placed in the middle part on the left side, and the context menu directly next to the selected object that allows to select properties. 
User feedback further revealed that the context menus were appreciated for offering quick access to common actions (e.g., duplicate, delete, change the z-index, modify object properties). The most liked functionalities were the ability to duplicate objects, to make multiple selections, to position objects in different layers, and to modify the background color. As compared to the first user study conducted with Figma and its right-anchored menu, the physical fatigue was considerably reduced. % 

Despite this general improvement over Figma, the second user study also showed that some issues still remain with Miro, such as the position of the zoom widget placed in the bottom right, which was disliked by all participants.

In addition, the second user study has shown that the interaction modality and the WSD configuration has impacts onto the experience of the user, which we will discuss further in the following sections.

\subsection{Benefits and issues with each modality}
%interaction modalities
    %normal interaction techniques
All three interaction modalities used in the user studies were deemed familiar by both p1 and p2, which made them easier to interact with the WSDs (``\emph{It resembles the interaction modalities that we know on the desktop. It embraces the great metaphors we are accustomed to.}'' p1, ``\emph{It uses standard interaction methods.}'' p2). 
However, we found that in each of the conditions, participants faced issues and appreciated other features. Based on our observations, we categorized them into pros and cons, described below and summarized in Table~\ref{tab:observationsXP2}. 

\begin{table*}[]
\centering
\caption{Pros and cons with each interaction modality used to design at scale 1:1 on a WSD.}
\label{tab:observationsXP2}
\begin{tabular}{llll}
         & \textbf{Keyboard \& Touchpad} 
         & \textbf{Touch}                       & \textbf{Tablet}                                                                                                                                                                                 \\
\hline
Benefits 
& \begin{tabular}[t]{@{}l@{}}- Precise positioning (using arrow keys)\\ - Easier text entering\\ - Use of shortcuts\\ - Lighter than a tablet\end{tabular} 
& \begin{tabular}[t]{@{}l@{}}- Better sense of space\\ - Instinctive, direct, and immediate\end{tabular}                                                                & \begin{tabular}[t]{@{}l@{}}- Dissociated viewports of tablet and WSD,\\ allowing two distinct views,\\ - Easier work from a distance, reducing physical \\ fatigue as it is possible to sit down\\ -  Easier work on tiny objects \end{tabular}                                                                                                                                                  \\
\hline
Issues   
& \begin{tabular}[t]{@{}l@{}}- Touchpad not appreciated\\ - Not comfortable and wrist pain\\ - Workspace moves a lot\\ - Unintended action triggering\end{tabular}     
& \begin{tabular}[t]{@{}l@{}}- Physical fatigue\\ - Arm and neck pain\\ - Not so robust\\ - Unintended action triggering\\ - Not so precise (tiny objects cannot be done)\\ - Requires a lot of zoom and pan\end{tabular} 
& \begin{tabular}[t]{@{}l@{}}- Arm and neck pain\\ - Weight of the tablet\\ - Fat finger effect\\ - Lag between the two views\\ - Working surface is small\\ - Long-range drag and drop is difficult\\ - Virtual keyboard not appreciated\end{tabular}
\\
\hline
\end{tabular}
\end{table*}

    %prefered modalities
P1 preferred the touch modality for the sense of acting directly on the WSD, then placed the tablet as second choice, with the keyboard last (``\emph{My favorite modality is the touch, although I experienced some glitches that caused problems.}'' p1, ``\emph{With the keyboard, positioning the objects is more precise and entering text is easier, but with the tablet I have better control on the zoom level on the display, and no unwanted zoom.}'' p1). P2 preferred the tablet, then ranked the keyboard second owing to the ability to use shortcuts, and ranked the touch modality last.
Touch was the most tiring interaction modality for both participants, because it required more movement.
Concerning the keyboard, participants would have liked to use it with a mouse rather than a touchpad. As for the tablet, participants would have preferred a larger tablet, and would have liked to use a stylus as well as a physical keyboard.

    %participants strategies with devices
When an interaction device was used in the given condition, p1 generally preferred to leave it on the table. When movement was necessary, he preferred to move the table, e.g., when accessing the zoom menu at the right bottom on WSD-VW (``\emph{I preferred not to carry the keyboard in hand for comfort. I preferred to move the table, that was less painful.}'' p1). In such situations, p2 preferred instead to carry the device around for more flexibility and to avoid wrist pain linked to entering text at an uncomfortable angle (``\emph{I carried the keyboard with me in order to be more mobile, more flexible and to avoid wrist pain.}'' p2, ``\emph{Having the tablet on the table was eventually an uncomfortable position for the neck, so, in the end, I preferred to hold the tablet in my hand.}'' p2). Ultimately, both users complained of physical fatigue and that the positioning of the interaction devices was a pain-inducing issue.
Interaction devices also divided the participants' attention between the device and the WSD, typically when entering text (``\emph{[I] divided my attention between the display and the keyboard.}'' p1).

Our results also show the main issues related to each interaction modality. 
In the keyboard + touchpad setting, the touchpad was too sensitive, which resulted in erratic movements of the workspace and created frustration. Also, typing text induced wrist pain due to the uncomfortable typing position, but this may be solved with a tilting table. Direct touch on the WSD induces physical fatigue and pain in the arm and neck. It lacks precision, particularly in the manipulation of tiny objects, and requires a lot of zoom and pan interactions. Some issues may be related to the touch hardware we are using, as it lacked robustness according to the participants. Finally, the tablet weight induces pain in the arm and neck, that could be alleviated by a tilting table, the working surface is too small, and a fat finger effect was observed, the long-range drag and drop is difficult, and lastly, there is a synchronization latency between the tablet and the WSD.
    %precision is an issue
%The participants noted that drawing a straight line was not easy with the tablet and touch conditions.

    %conclusion ; mix of techniques
Overall, the participants would have preferred to have the ability to choose and switch between several interaction methods throughout the design session.

\subsection{Difference between both WSDs}
Our observations also indicate some differences between the two WSD settings. 
%differences between both WSDs
    %bezel
First, bezel issues were raised more frequently with WSD-IA than WSD-VW, likely because they are wider on WSD-IA, even if they are more numerous on WSD-VW.

    %zoom
The position of the zoom widget was disliked by participants on WSD-VW, although they did not comment on it on WSD-IA. 
This is probably due to the fact that it was displayed in the back of the participants on WSD-IA, and went unnoticed. Only p2 used these options on WSD-IA and only once, with the keyboard modality.

    %physical ergonomics
The physical ergonomics play an important role and must be handled with care.
        %table
The height-adjustable and movable table in WSD-VW was liked and used, even if it was not ergonomic enough. Indeed, the participants were quite tall and the maximum height of the table in WSD-VW was insufficient for them. P2 adapted the table height in WSD-IA for more comfort. Participants wished they could tilt the table, especially with the keyboard and the tablet conditions to avoid wrist, shoulder, neck and back pain.
        %chair
P2 sometimes sat down in WSD-VW (``\emph{I was not too tired, but maybe it was because I sat down.}'' p2 when using the keyboard on WSD-VW). But, none of the participants sat in WSD-IA, likely because available chairs were behind the screens, i.e., outside the participants' field of view. It was noted by p2 that being able to sit down in WSD-VW was less tiring when using the keyboard and the tablet, and offered better positioning for entering text.

        %space around
Finally, the space around WSD-VW was appreciated (``\emph{It was nicer with [WSD-VW] than with [WSD-IA] because there was more space, and I sat down. I was able to arrange my desk better.}'' p2) and led to a better user experience, likely because participants realized they could use a chair and sit down. %Even if only p2 sat down.

\subsection{Implications for design}\label{sec:guidelines}
Based on our results from the two user studies, we can propose several \textbf{design guidelines} for creating a WSD design tool: 
\begin{itemize}
    \item Main actions should be reachable from the work area (e.g., creation of a new object, selected object properties, deletion, layering management). %Have a main menu that can be reached from the work area (avoid placing it at the top or the bottom).
        %\item Make the properties of the currently selected object reachable from the work area
    %\item Create a list of created objects and make it reachable from the work area
    \item Dialogue boxes should be avoided, or opened near the work area.
    \item All design tool elements should be legible from near and far, and actionable with the used modality. %Have a sufficiently big font size for the design tool interface, or the possibility to zoom in and out + buttons etc.
    \item It should be possible to use the entire WSD surface for design and that the design tool elements do not get in the way, e.g., by making the design tool menus retractable, or movable. All context menus should be movable to avoid the bezels. %i.e., menu, list of created objects, and properties of the currently selected object.
    \item A distinct device with a screen, like a tablet, should provide a second view for zooming in and carrying out design actions.
    %Support a solution to access the workspace from a distinct display device like a tablet
    \item It should be possible to lock the viewport of the workspace, with independent settings for zoom and pan actions.
    \item A possibility for resetting the viewport to a defined setting should be provided, as reopening the design at the same position.
    \item The system should support very large workspaces (gigapixel resolution).
    \item It should be possible to change the background color.
    \item It should be possible to duplicate (multiple) objects.
    \item Expert interactions, e.g., shortcuts, should be supported.
    \item There should be an enriched objects’ library with advanced widgets, graphs, and icons to choose from.
\end{itemize}

Overall, regarding the different interaction methods, and based on our results, the best approach seems to be to \textbf{provide a mix of modalities}, including touch interaction for direct manipulation of objects on the WSD, a distinct device with a dissociated view (e.g., tablet), and a physical keyboard as efficient means to enter text.

% \end{itemize}

Regarding \textbf{the design of the WSD itself}, it is better to reduce as much as possible the bezels between the display tiles. It is also important to ensure the visibility of the cursor at all times, and to facilitate long-range interactions, e.g., through pointing facilitation techniques~\cite{BALAKRISHNAN2004857}.

Flat and circular displays have their pros and cons. 
Circular displays reduce the amplitude and the amount of movement required, text is easier to read from everywhere, but there is always a part of the screen at the back that can be invisible or forgotten, and there is less space to add furniture to make the interaction more comfortable. 
Flat displays, i.e., the most common form of WSDs, require much physical navigation. The user must move a lot to see the content on the other side of the screen and to get an overview. But there is more space around the display, which makes it easier to adapt it for a better user experience, for instance, by adding furniture.

%maybe different interaction methods and/or functionnalities depending of the design process stage. Indeed the needs are not the same during the first sketching, the wireframeing, the advanced mockup and the prototyping. 

To improve the user experience regarding the \textbf{space around the WSD}, we propose to provide:
%\begin{itemize}
    %\item 
    a height-adjustable mobile and tilting table to park interaction devices, and
    %\item 
    chairs for users to rest.
    %\item 
   
%\end{itemize}

\subsection{Limitations}
%\color{blue}
Concerning the user studies, our findings come with several limitations. 
First, the task was limited to the reproduction of an existing UI prototype for a WSD environment, and no creative part was involved. 
In addition, the user studies involved only three participants who were all unfamiliar with the design tools used. The advantage here, however, was that they had no prior habits, e.g., using specific shortcuts, and were not frustrated by not being able to work as quickly as an expert would on a familiar software. 

%\color{red}
%% LEARNING EFFECT
In the second user study, a clear learning effect was observed between the first (played on WSD-IA) and last sessions (played on WSD-VW). Indeed, participants were not experts at using Miro, despite having used it for other purposes than interface design. They discovered some features and faster ways of performing some actions during the sessions. This makes it difficult to compare the completion levels between them.

Finally, it must be mentioned that we encountered (unexplained) bugs with the display of the context menu in Miro, %maybe due to the high resolution used, 
and that bezels would sometimes occlude menu options. Also, the mock-up covered mainly the left part of the display.
%Participants did not criticize the left and centered menu to create objects, suggesting that this solution seems better adapted to the WSD context than the top menu in Figma. 
We expect more issues to arise when users need to create objects on the right side of the display using touch (particularly for flat WSD such as WSD-VW) or the keyboard interaction methods.

%\color{black}
\section{Conclusion and Future Work}\label{sec:conclusion}
In this paper, we presented two user studies and a comparative analysis investigating the suitability of existing UI design tools to design at 1:1 scale on WSDs using three different interaction methods: touch, a keyboard with a touchpad, and a tablet. 
Our main takeaways are that 
\begin{enumerate*}[label=(\roman*)]
    \item prototyping at 1:1 scale and being able to see live the final rendering in real time is appreciated,
    \item tablet-based interaction proved to be the most comfortable,
    \item each interaction method has its benefits and drawbacks; %\color{red}
    using a mix of modalities is promising,
    \item the spatial positioning of interaction elements on the screen, as well as the design of the physical environment, is of utmost importance to reduce physical fatigue and support efficient work,
    \item %\color{red}
    among the nineteen desktop-optimized design tools analyzed, none satisfies all the criteria, but Miro is the most suitable.
\end{enumerate*}

%\color{red}
So, to answer our research questions, a desktop optimized tool cannot be used as-is to design at 1:1 scale in a WSD environment. To improve the design of such a tool, we propose 12 design recommendations indicating how UI elements should be placed, and which features need to be provided, see Section~\ref{sec:guidelines}.
Regarding interaction modalities, the best approach for this type of design currently seems to be the synchronized tablet, but the other modalities also provide some unique benefits. A mix of interaction methods seems promising and is yet to be tested. 

However, these studies involved few users and are focussed on the reproduction of a UI design. To validate the proposed guidelines, further experiments are needed. 

%In the future, we will reiterate the study to expand the number of participants to verify our initial observations, while making sure the ordering of conditions mitigates learning effects. We will also explore other prototyping tools.
More specifically, future work needs to further explore how 1:1 scale design can be supported by focussing on creating a new design rather than reproducing an existing one.
As design is often a collaborative task, future research also needs to investigate how to manage collaboration during design sessions at 1:1 scale on WSDs. 
Additional questions are how modalities can be mixed and support the transitioning between multiple interaction methods in the same session.
The studies from this paper enabled us to better understand the requirements for technical systems and the related problems and opportunities when using them to design for WSDs. In a longer perspective, this work is done as part of a design science methodology~\cite{hevner2007three}, ultimately aiming to create a design tool that is suitable for WSDs. The results presented in this paper are part of the first step, the relevance cycle, and will enable us to identify the user requirements.

\section{Acknowledgments}
We thank all user study participants.
%\color{black}

% % ======== References =========
% \begingroup
% \sloppy
% \printbibliography
% \endgroup 

%=====================================
\bibliographystyle{IEEEtran}


%\bibliography{biblio}

% \bibitem{b1} G. Eason, B. Noble, and I. N. Sneddon, ``On certain integrals of Lipschitz-Hankel type involving products of Bessel functions,'' Phil. Trans. Roy. Soc. London, vol. A247, pp. 529--551, April 1955.
% \bibitem{b2} J. Clerk Maxwell, A Treatise on Electricity and Magnetism, 3rd ed., vol. 2. Oxford: Clarendon, 1892, pp.68--73.
% \bibitem{b3} I. S. Jacobs and C. P. Bean, ``Fine particles, thin films and exchange anisotropy,'' in Magnetism, vol. III, G. T. Rado and H. Suhl, Eds. New York: Academic, 1963, pp. 271--350.
% \bibitem{b4} K. Elissa, ``Title of paper if known,'' unpublished.
% \bibitem{b5} R. Nicole, ``Title of paper with only first word capitalized,'' J. Name Stand. Abbrev., in press.
% \bibitem{b6} Y. Yorozu, M. Hirano, K. Oka, and Y. Tagawa, ``Electron spectroscopy studies on magneto-optical media and plastic substrate interface,'' IEEE Transl. J. Magn. Japan, vol. 2, pp. 740--741, August 1987 [Digests 9th Annual Conf. Magnetics Japan, p. 301, 1982].
% \bibitem{b7} M. Young, The Technical Writer's Handbook. Mill Valley, CA: University Science, 1989.
% \end{thebibliography}
% \vspace{12pt}
% \color{red}
% IEEE conference templates contain guidance text for composing and formatting conference papers. Please ensure that all template text is removed from your conference paper prior to submission to the conference. Failure to remove the template text from your paper may result in your paper not being published.

\end{document}